\documentclass[]{mn2e}

\newif\ifAMStwofonts

\usepackage{graphicx}

\def\bib{\parskip=0pt\par\noindent\hangindent\parindent \parskip =2ex plus .5ex
    minus .1ex}
\newcommand{\lsim}{\raisebox{-0.13cm}{~\shortstack{$<$ \\[-0.07cm] $\sim$}}~}
\newcommand{\gsim}{\raisebox{-0.13cm}{~\shortstack{$>$ \\[-0.07cm] $\sim$}}~}

\title[The evolution of K$_s$-selected galaxies]{The evolution of K$\rm _s$-selected galaxies in the GOODS/CDFS deep ISAAC field}

\author[K.I. Caputi, J.S. Dunlop, R.J. McLure and N.D. Roche] {K.I. Caputi\thanks{Present address: Institut d'Astrophysique Spatiale,
B\^at. 121, Universit\'e Paris XI, 91405 Orsay Cedex, France. \newline
E-mail: kcaputi@ias.u-psud.fr}, J.S. Dunlop, R.J. McLure and 
N.D. Roche \thanks{Present address: Instituto de Astronom\'{\i}a, Universidad Nacional Aut\'onoma de M\'exico, PO Box 439027, San Diego, CA 92143-9027}
       \\ Institute for Astronomy, University of Edinburgh, Royal Observatory,
       Edinburgh EH9 3HJ, Scotland, U.K. \\ }

\date{ }

\pubyear{2005}

\begin{document}

\maketitle

\label{firstpage}

\begin{abstract}

  We present estimated redshifts and derived properties for a sample of 1663 galaxies with $\rm K_s \le 22$ (Vega), selected from 50.4 $\rm arcmin^2$ of the GOODS/CDFS field with deep ISAAC imaging, and make an extensive comparison of their properties with those of the extremely red galaxies (ERGs) selected in the same field. We study in detail the evolution of $\rm K_s$-selected galaxies up to redshifts $z \sim 4$ and  clarify the role of ERGs within the total $\rm K_s$-band galaxy population.  We compute the total $\rm K_s$-band luminosity function (LF) and compare its evolution with the ERG LF. Up to  $\langle z_{phot}\rangle=2.5$, the bright end of the $\rm K_s$-band LF shows no sign of decline, and is progressively well reproduced by the ERGs with increasing redshift. We also explore the evolution of massive systems present in our sample:  up to 20\%-25\% of the population of local galaxies with assembled stellar mass $\rm M>1 \times 10^{11} M_\odot$  has been formed before redshift $z \sim 4$, and contains $\sim$ 45\% to 70\% of the stellar mass density of the Universe at that redshift. Within our sample, the comoving number density of these massive systems is then essentially constant down to redshift $z \sim 1.5$, by which point most of them  have apparently evolved into $\rm (I-K_s)$-selected ERGs.  The remaining massive systems observed in the local Universe are assembled later, at redshifts $z \lsim 1.5$. Our results therefore suggest a two-fold assembly history for massive galaxies, in which galaxy/star formation proceeds very efficiently in high mass haloes at very high redshift.

\end{abstract}


\setcounter{figure}{0}

\section{Introduction}
\label{sec-intro}

\parskip=0pt
 
     The study of near-infrared (near-IR) galaxies, in particular K-band selected galaxies, has increasingly been recognised as an efficient method for setting constraints on the formation epoch of massive galaxies.   The rest-frame K-band is the most suitable photometric tracer of stellar mass, relatively independent of the galaxy star formation history. Thus, observer-frame K-band selected samples are expected to contain all of the most massive galaxies at least up to redshifts $z \sim 1-2$. Based on the analysis of near-IR-selected galaxy samples, different authors have studied the evolution of the stellar mass content of the Universe up to redshift $z \sim 2$ (Glazebrook et al. 2004, Drory et al. 2004, Fontana et al. 2004), and in small-area surveys up to redshift $z \sim 3$ (Fontana et al. 2003, Dickinson et al. 2003, Rudnick et al. 2003).  These studies have provided evidence for the existence of massive systems up to at least these high redshifts.

     Among the K-band selected sources, extremely red galaxies (ERGs) are commonly considered to trace the progenitors of the local early-type population at high redshift. In particular, in a bright sample of ERGs, Daddi et al. (2000) determined the existence of strong  clustering, a characteristic property of local elliptical galaxies. This strong clustering was later confirmed to extend to fainter magnitudes  by Roche et al. (2002) and  Roche, Dunlop \& Almaini (2003) (hereafter R03). However, it is not yet clear whether the ERG population contains  {\em all} the progenitors of the local early-type galaxies, or whether a total K-band sample of galaxies is necessary to select all of these progenitors at high redshift. To investigate the evolution of K-band selected galaxies and to assess the role of ERGs within the total K-band galaxy population, we present here the study of a sample of 1663 galaxies with $\rm K_s \leq 22$ (Vega), and compare their properties with those obtained for a sample of 198 ERGs selected in the same field. The selection of the ERG and the total $\rm K_s$-band samples from the same field is essential to accurately compare the properties of both populations, by minimising the potential existence of systematic errors due to cosmic variance.

\begin{figure}
\begin{center}
\includegraphics[width=1.0\hsize,height=0.95\hsize,angle=0] {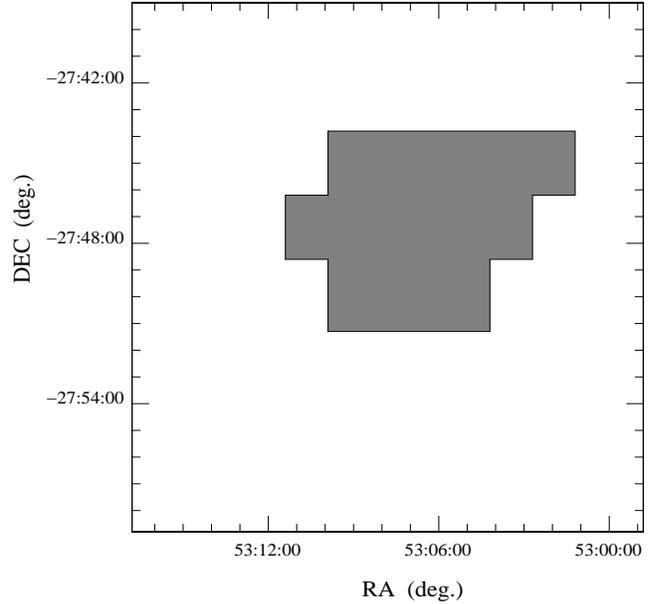}
\caption[]{\label{isaac} Schematic diagram of the sub-region covered by deep
VLT/ISAAC near-infrared imaging in the GOODS/EIS v0.5 release (shaded area), within the larger GOODS/CDFS field
with full HST-ACS coverage.}
\end{center}  
\end{figure}

The Great Observatories Origins Deep Survey (GOODS) (Giavalisco et al. 2004) is
providing unprecedented multiwavelength data in $\sim$ 320 arcmin$^2$ centred on
the Chandra Deep Field South (CDFS) and Hubble Deep Field North (HDFN). Within
the GOODS/CDFS field, R03 selected 198 ERGs with $\rm K_s \leq 22$ (Vega) and $\rm (I_{775}-K_s)>3.92$ from 50.4 arcmin$^2$ for which deep near-IR data have been obtained with the Infrared Spectrometer and Array Camera (ISAAC) on the `Antu' Very Large Telescope (Antu-VLT) (GOODS/EIS v0.5 release). This is the deepest significant sample of ERGs selected to date. The region covered by the ISAAC v0.5 imaging within the GOODS/CDFS field is shown in Figure \ref{isaac}. The redshift distribution and other derived properties of the R03 ERG sample have been studied by Caputi et al. (2004); hereafter C04. Both R03 and C04  are complementary to other previous studies of shallower samples of ERGs in the CDFS (Cimatti et al. 2002, Moustakas et al. 2004). 
C04 constructed the ERG luminosity function (LF) and determined that its bright end does not evolve from redshifts $\langle z_{phot}\rangle=2.0$ to $\langle z_{phot}\rangle=2.5$, connecting this fact with the presence of progenitors of local $\rm L>L^{\ast}$ galaxies at redshifts $z_{phot}>2$.  Also, C04 showed that the comoving densities of ERG progenitors of local $\rm L>L^{\ast}$ galaxies at different redshifts are below the total expected values, indicating either that ERGs  cannot account for all the progenitors of massive galaxies at redshift $z=0$, or that the underlying assumption of passive evolution is not completely valid.

  Within the GOODS program, near and mid-IR data from the Spitzer Space Telescope will soon become available for the GOODS/CDFS. Spitzer is currently imaging this field with the Infrared Array Camera (IRAC) at $3.6 \mu m$, $4.5 \mu m$, $5.8 \mu m$ and $8.0 \mu m$, and with the Multiband Imaging Photometer (MIPS) at $24 \mu m$. The availability of these data will provide a unique opportunity to follow up ISAAC and HST sources in the CDFS,  and observe directly at wavelengths which map the rest-frame $\rm K_s$-band at high redshift.  Also, the existence of mid-IR images will  allow the study of the light re-emitted by dust in optical and near-IR sources. The analysis of mid-IR images will be an important tool to understand, for example,  the extent to which the presence of dust is responsible for the extremely red colours observed in ERGs.

In the present work we use the public reduced GOODS/EIS v0.5 data to select a full sample of galaxies with $\rm K_s \le 22$ (Vega) in the same area analysed by R03. The aim is to study the evolution of  the total $\rm K_s$-galaxy population with redshift and to compare the results with those obtained for the R03 ERG sample by C04.  The layout of this paper is as follows. First, in Section 2, we give details on the sample selection and the multiwavelength photometry. In Section 3 we explain the photometric redshift techniques applied, and discuss the calibration of the redshift estimates with spectroscopic data and with the previous estimates for ERGs determined by C04. In Section 4 we present our results, and compare with the results for ERGs obtained by C04:  we discuss raw and dust-corrected Hubble diagrams, the evolution of the LF, and the evolution of massive systems up to redshift $z_{phot}=4$. In Section 5, we predict IR magnitudes and colours for our $\rm K_s$-selected galaxies in three Spitzer/IRAC channels: $3.6 \mu m$, $4.5 \mu m$ and $5.8 \mu m$. Finally, in Section 6 we offer some concluding remarks.  We adopt throughout a cosmology with $\rm H_o=70 \,{\rm km \, s^{-1} Mpc^{-1}}$, $\rm \Omega_M=0.3$ and $\rm
\Omega_\Lambda=0.7$.

\section{The sample}
\parskip=0pt

\subsection{Multiwavelength photometry}
 
GOODS observations include optical and near-IR imaging in the B, V, I$\rm
_{775}$ and z bands with the Advanced Camera for Surveys (ACS) on board the Hubble Space Telescope (HST), and J, H and $\rm K_s$ bands with the ISAAC/VLT. The public reduced ISAAC images (GOODS/EIS v0.5 release) consist of eight maps covering approximately one third of the whole GOODS/CDFS area with ACS/HST coverage. We constructed our $\rm K_s$-band catalog using the reduced $\rm K_s$-band images and extracting sources with the public code SEXTRACTOR (Bertin \& Arnouts, 1996) on each map separately. As one of our aims was to compare our results with those obtained for ERGs by R03 and C04, we needed to select the $\rm K_s$-band sample in the same way and, consequently, we used the same SEXTRACTOR parameters used by R03 for the extraction of sources, i.e. a detection threshold of $1.4 \sigma_{sky}$ in at least six contiguous pixels. We employed a Gaussian kernel with a 3-pixel FWHM and the corresponding weighting maps. The low signal-to-noise edges of each map have been excluded from the detection procedure, leaving a total effective area for detections of 50.4 arcmin$^2$. There is some overlap between the eight $\rm K_s$-band maps (even after the trimming of the edges). Thus, as we ran SEXTRACTOR on each of the maps separately, we discarded repeated sources appearing on different map catalogs. We used total Kron-type magnitudes for limiting the sample to objects with magnitude $\rm K_s \leq 22$ (and above a $3\sigma$ detection limit, although only a few objects with $\rm K_s \leq 22$ were below this threshold). The number of sources in this first $\rm K_s=22$ catalog was 1756. For measuring colours and applying photometric redshift algorithms, we used circular 2-arcsec diameter aperture magnitudes. This aperture size is the same as used for the ERGs by R03 and C04. We ran SEXTRACTOR in `double-image mode' to perform aperture photometry on the J and H band images, using the corresponding $\rm K_s$-band images for the detection of sources.  For a minority of objects ($<$10\%), a 2-arcsec diameter aperture produced a difference $>$ 0.5 mag between the aperture and the total $\rm K_s$ magnitudes. For these objects we used a larger aperture of $2 \sqrt{2} \approx 2.83$-arcsec diameter. The aperture sizes we adopted for the near-IR photometry have been deliberately selected to coincide with aperture sizes used to construct the ACS catalogs, as we explain below.

 We looked for counterparts of our 1756 $\rm K_s$-band sources in the public r.1.0z GOODS/CDFS ACS catalogs, within a matching radius of 1 arcsec. The GOODS/ACS catalogs have been constructed running SEXTRACTOR on the version v1.0 of the reduced stacked  GOODS/CDFS ACS images. The extraction of sources has been made on the z-band images and SEXTRACTOR has been run in  `double-image mode' to perform photometry on the B, V and I$\rm_{775}$ bands. The GOODS/ACS catalogs provide photometry in 11 different circular apertures for each source. For each of our  $\rm K_s$-band counterparts, we used the magnitudes corresponding to the same aperture size we had selected in the near-IR bands. We found that 25/1756 of our K$_s \le 22$ sources did not have a counterpart in the GOODS/CDFS ACS catalog. We made an individual inspection of each of these 25 sources on the $\rm K_s$-band images and found that 13/25 were spurious $\rm K_s$-band detections, most of them close to the (already trimmed) edges of the corresponding $\rm K_s$-band map and, in some cases, with no flux in the J or H bands. Another one of the  25 sources appears in a region where the ACS images are damaged by satellite tracks, so all the optical information is missed for this object. After exclusion of these 14 sources, the remaining sample had 1742 objects.  In each case of non-detected flux (SEXTRACTOR magnitude 99.0) in any optical or near-IR band, we measured the flux manually in the corresponding aperture using the IRAF task `phot' on the  stacked GOODS HST/ACS V1.0 and the VLT/ISAAC v0.5 images. For faint objects in every ACS band (all of them ERGs), we repeated the same procedure explained in C04, leaving only a few sources as formally non-detected in the V or redder bands in the multiwavelength catalogs we used as input for the photometric redshift algorithms.

  As a final comment on the photometry of $\rm K_s$-selected galaxies, we note that we used the photometric measurements made on the public GOODS Hubble Ultra Deep Field (HUDF) images in the B, V, I$\rm_{775}$ and z bands for three ERGs. These objects were initially recognised as possible  $z>4$ candidates, but the better quality of the HUDF data allowed to reveal them as being at lower redshifts (C04).

\subsection{Completeness}

 We used the IRAF tasks `gallist' and `mkobjects' to create a $\rm K_s \leq 22$ mock catalog of 200 objects with a power-law luminosity distribution and insert them into each of the eight ISAAC $\rm K_s$-band maps. We ran SEXTRACTOR again on these images, using the same original extraction parameters, and checked the fraction of artificial sources recovered in each case. From this, we estimated that our sample was  100\%, 89\%, 86\%, and 80\% complete at $\rm K_s \leq 20.0$, $\rm 20.0 < K_s \leq 21.0$, $\rm 21.0 < K_s \leq 21.5$ and $\rm 21.5 < K_s \leq 22.0$, respectively.

\subsection{The ERG sample}
 
    In this work, we compare the redshift distribution and derived properties for  $\rm K_s \le 22$-selected galaxies with the corresponding results obtained by C04 for the ERGs selected by R03 in the same field.  However, in order to assess the validity of the comparison of the present $\rm K_s$-selected sample with a subsample of ERGs obtained in an independent way,  we checked the number of objects with $\rm (I_{775}-K_s) > 3.92$ among the new $\rm K_s \leq 22$ sources, obtaining a number of 205. 186/205 of these objects are in common with the sources selected by R03 as ERGs with the same colour cutoff (which are 198 objects in total), producing an overlap of both ERG samples $>$90 \%. Thus, the comparison of results for the present $\rm K_s$-selected sample and the R03 ERGs is entirely valid.

\subsection{Star/galaxy separation}
 
  We cleaned our catalogs for stars/QSOs using the SEXTRACTOR stellarity parameter  `CLASS\_STAR', as  measured on the ACS z-band images and appears on the GOODS/CDFS ACS catalogs. We found that 107/1742 $\rm K_s \leq 22$ sources had z-band counterparts with CLASS\_STAR$\geq 0.8$. 64 out of these 107 objects are confirmed by the SIMBAD astronomical database as stars or QSOs, so we removed them from our catalogs, leaving 1678 sources for the photometric redshift algorithm input files. The inspection of the redshift output catalogs showed that 15 out of the remaining 43 objects with CLASS\_STAR$\ge 0.8$ could not be satisfactorily modelled by any galaxy spectral energy distribution (SED) template (cf. Section 3) and had zero probability of being at any redshift. Thus, we considered that these 15 objects  were also very likely to be stars/QSOs and excluded them from our sample, leaving 1663 sources in the final  $\rm K_s \leq 22$ galaxy catalog.

\subsection{Number counts}
  
\begin{figure}
\begin{center}
\includegraphics[width=1.0\hsize,height=0.95\hsize,angle=0] {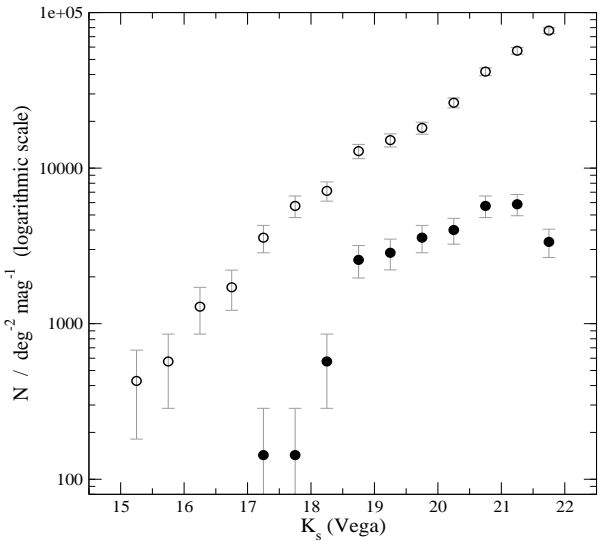}
\caption[]{\label{numcounts} Differential number counts for our $\rm K_s$-selected galaxies (empty circles), in comparison with the number counts for the ERGs selected in the same field by R03 (filled circles). }
\end{center}  
\end{figure}

 Figure \ref{numcounts} shows the differential number counts for our sample of $\rm K_s$-selected galaxies (empty circles) in comparison with the number counts for ERGs obtained by R03 (filled circles). Both sets of counts have been corrected for incompleteness. The ERG number-count curve is characterised by a flattening of the count slope beyond $\rm K_s \approx 19.5$. In Section 4.3, we show that this fact is related to the existence of a turnover in the ERG LF at redshift $z_{phot} \approx 1$.

\section{Redshift estimates}

\subsection{Photometric redshift techniques}

  We used the public code HYPERZ (Bolzonella, Miralles \& Pell\'{o}, 2000) to compute photometric redshifts ($z_{phot}$) for our sample of $\rm K_s$-selected sources, using the seven passbands described in Section 2.1 (B, V, I$\rm_{775}$, z, J, H and K$\rm_s$) and the GISSEL98 SED template library (Bruzual \& Charlot 1993). A detailed explanation of the HYPERZ performance has been given in C04. We only note here that we refer to the HYPERZ `primary solution' as the redshift estimate $z_{phot}$ (and corresponding set of parameters) which produce the maximum likelihood in the SED fitting. The `secondary solution' is the second most likely solution in parameter space. We ran HYPERZ using the same parameter values as in C04. In particular, we applied a Calzetti et al. (2000) reddening law with the V-band extinction ($A_V$) varying between 0 and 1. For those objects taking the maximum possible reddening, we made a second HYPERZ run allowing  the V-band extinction $A_V$  to vary from 0 to 3. We used the public `Bayesian photometric redshift' (BPZ) code by Ben\'{i}tez (2000) to obtain a second, independent redshift estimate for a subset of the sources that HYPERZ found to be at high redshift, namely:
\begin{itemize} 
\item very bright $\rm K_s$-band sources with primary solution $z_{phot}>2$ 
\item sources with significant flux in the B-band ($\rm B_{AB} \leq 27.5$) and primary solution  $z_{phot}>3$ 
\item all sources with primary solution $z_{phot}>4$. 
\end{itemize}

\noindent For those objects which BPZ also placed at high redshift, we accepted the HYPERZ primary solution as the definitive solution for the object. On the other hand, when BPZ determined a lower redshift estimate, we adopted the HYPERZ secondary solution, which was in agreement with BPZ in most of these alternative cases. For seven sources, we found that neither  the HYPERZ primary nor secondary solutions provided as low a redshift estimate as the BPZ determination. In these cases, we adopted the BPZ estimates as the definitive redshifts. One extra source had a very high redshift HYPERZ primary solution and a low redshift secondary solution in agreement with BPZ. However, the real redshift for this source  is $z_{phot}=1.669$, based on spectroscopic determinations (cf. Section 3.2), which both photometric redshift algorithms fail to correctly estimate. In this case only, we adopt the spectroscopic value as the definitive redshift. For the eight sources with either BPZ or spectroscopic redshift replacements, we do not have the complementary information provided by HYPERZ in the output, i.e. SED best-fit template, k-corrected $\rm K_s$-band absolute magnitude, age, etc.

  As we explained in Section 2.1, after running HYPERZ on our catalog of 1678  $\rm K_s \leq 22$ sources, we determined that 15 of these objects were likely to be stars/QSOs, based on their stellarity parameter and the fact that they did not have a satisfactory fitting by any of the templates in the HYPERZ SED library. After rejecting these objects, our final $\rm K_s \leq 22$ catalog consisted of 1663 galaxies, including $\sim$ 95 \% of the ERGs in the R03 sample. (The few ERGs mentioned in C04 as likely stellar contaminants are now automatically rejected from the present $\rm K_s \leq 22$ galaxy sample).

\begin{figure}
\begin{center}
\includegraphics[width=1.0\hsize,angle=0] {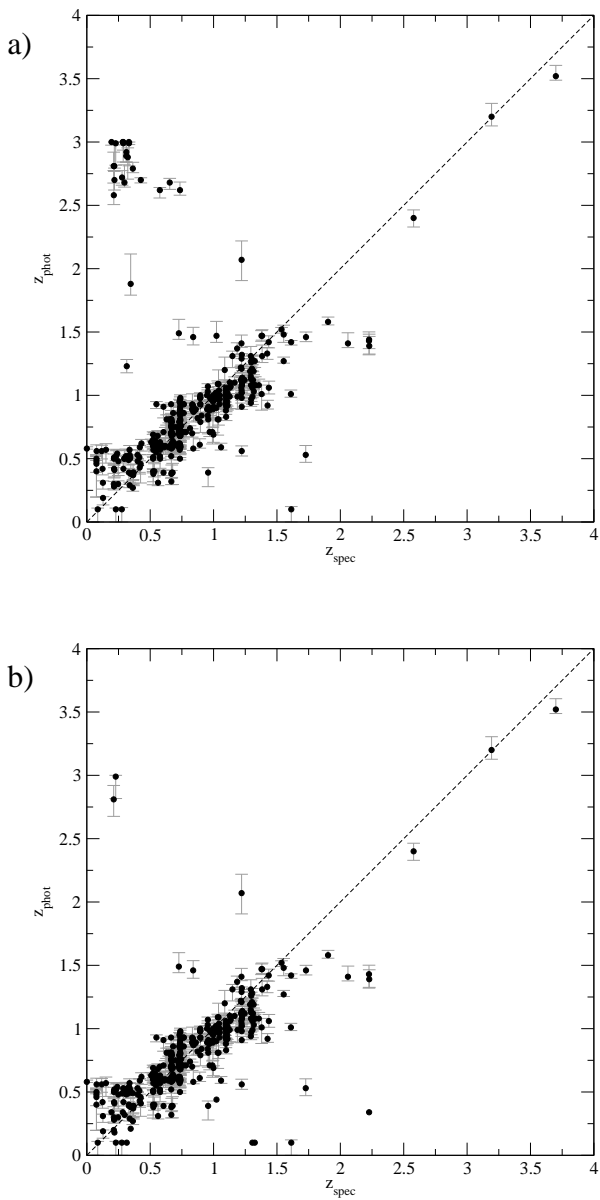}
\caption[]{\label{zvsspecz} Photometric redshifts vs. spectroscopic redshifts for $\rm K_s$$\le$22-selected galaxies in the GOODS/CDFS deep ISAAC field: a) before the calibration; b) after the calibration. }
\end{center}  
\end{figure}

\subsection{Photometric redshift calibration}

  We used the public spectroscopic redshifts available for the GOODS/CDFS to calibrate our photometric redshift estimates. Spectroscopic data in the GOODS/CDFS have been obtained by different authors (Cimatti et al. 2002b, Vanzella et al. 2004, Le F\`evre et al. 2004, Szokoly et al. 2004, among others). All the spectroscopic redshifts available for sources in the r.1.0z GOODS/CDFS ACS catalog have been compiled in a master list at http://www.eso.org/science/goods/spectroscopy/
    \newline CDFS$\_$Mastercat. The spectroscopic redshifts obtained as part of the K20 survey (Cimatti et al. 2002b) have recently made public at http://www.arcetri.astro.it/$\sim$k20/spe\_release\_dec04/index.
\newline
  html.

  Figure \ref{zvsspecz}(a) compares our redshift estimates with the spectroscopic redshifts for the $\rm K_s \leq 22$ sources included in different CDFS spectroscopy samples. We restrict the comparison to sources with good quality flags for the spectroscopic redshift determinations. The error bars in the photometric redshifts correspond to 1$\sigma$-confidence levels. We observe  quite  good agreement between the redshift estimates and the real redshifts in most cases. However, there is an excess of low redshift sources which we identified as being at redshifts $z_{phot}>2.5$. These sources with mismatching results are indicating that the photometric redshift algorithms might have a tendency to favour high redshift values when degenerate solutions in parameter space exist.  To compensate for this effect, we calibrate our photometric redshifts with the following criteria: for sources with significant B-band flux ($\rm B_{AB} \leq 27.5$) and a large difference between the HYPERZ primary and secondary solutions, i.e. $ (z_{prim}-z_{sec})\geq 1$, we adopted the HYPERZ secondary lower-redshift solution in the cases in which this redshift had at least 40\% of the probability of the primary one. For sources with $\rm B_{AB}> 27.5$ and $(z_{prim}-z_{sec})\geq 1$, which are more likely to be at high redshift, we adopted a threshold of 70\% between the relative probabilities of the  secondary and the primary solutions to adopt the former as the preferred answer. These criteria have been applied in addition to those explained in Section 3.1, without affecting the cases in which the HYPERZ secondary solution was already selected for providing a better agreement with the corresponding BPZ estimation. In summary, we adopted the HYPERZ secondary solution for 136/1663 sources, while 1519/1663 remained with the primary one (and 8/1663 had either BPZ or spectroscopic value replacements). The comparison of our photometric estimates with the spectroscopic redshifts after the calibration procedure is shown in Figure \ref{zvsspecz}(b). The median of the relative errors between our estimates and the real redshifts is $|z_{phot} - z_{spec}|/ (1+z_{spec})=0.05$.   In particular, we obtain a median of $|z_{phot} - z_{spec}|/ (1+z_{spec})= 0.07$ for the compared sources with  $z_{phot}>1.5$. These values are among the typical uncertainties obtained in photometric redshift estimates available in the literature.

  Optical U-band data in the CDFS has been obtained by Arnouts et al.(2001). Given the shallower depth and poorer resolution ($1.5^{\prime\prime}$) of the U-band images in comparison with those corresponding to the GOODS/HST-ACS, we decided not to include the former  in the input for the photometric redshift algorithms. However, for consistency, we checked for the presence of counterparts of our high-redshift $\rm K_s$-selected galaxies among the CDFS U-band sources.  We found that, as expected, none of our   $\rm K_s$-selected galaxies with estimated redshift $z_{phot} \gsim 2.8$ has a counterpart in the CDFS U-band source catalog within a matching radius of 1 arcsec. Nevertheless, it should be kept in mind that fainter U-band counterparts could exist in some cases. Thus, the lack of deep U-band data could potentially produce that a fraction of our high-redshift ($z_{phot} \gsim 2.8$) sample  is contaminated with faint low-redshift galaxies with degenerate solutions in photometric redshift space.

  On the other hand, while we were performing the present study, a new release (v1.0) of the GOODS/CDFS ISAAC data in the $\rm K_s$ and J bands has been made public. The new release includes the determination of a uniform photometric calibration across the entire GOODS/CDFS field. We revised the $\rm K_s$ and J-band  photometry of our $\rm K_s$-selected sample in the new ISAAC maps. After running HYPERZ and BPZ under the same conditions as explained in Section 3.1, we determined the existence of a basically negligible impact of any possible difference between the ISAAC v0.5 and v1.0 calibrations on our estimated redshifts.

\subsection{The ERG photometric redshifts}

   The calibration procedure explained in Section 3.2 was not implemented in C04 because  CDFS spectroscopic redshifts were not yet completely available and because the overlap with the R03 ERG sample was small. Thus, we need to verify that the  present selection of a $\rm K_s$-band sample and the calibration of the photometric estimates based on spectroscopic redshifts  still produces results in agreement with C04 for the ERG subsample.  
   
\begin{figure}
\begin{center}
\includegraphics[width=1.0\hsize,height=0.95\hsize,angle=0] {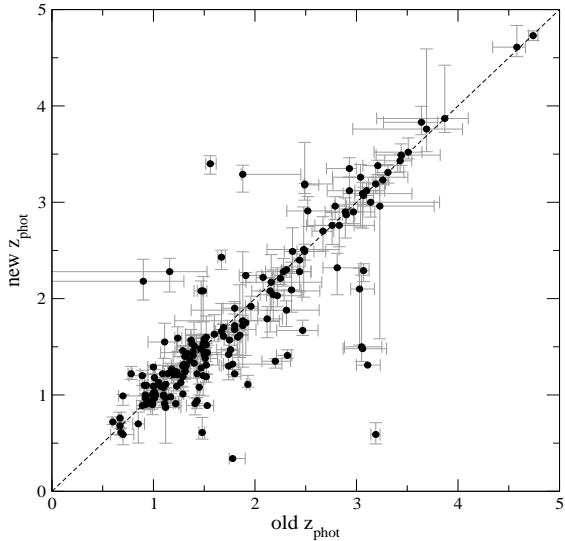}
\caption[]{\label{zvsz_eros} Redshift estimates determined in the present work for the R03 ERGs versus the redshift estimates obtained in C04.}
\end{center}  
\end{figure}

   Figure \ref{zvsz_eros} shows the new redshift estimates we determined for the ERGs versus the redshift estimates obtained by C04.  The error bars correspond to 1$\sigma$-confidence levels. Both sets of redshifts appear to be in very good agreement, except for a few outliers which correspond to cases of very strong degeneracies in parameter space. This correlation confirms that, in spite of the differences in the redshift calibration implemented in the present work, the redshift estimates  obtained for the ERGs are consistent and the comparison of derived properties for our total $\rm K_s$-selected sample with the results obtained in C04 for the ERGs is valid.

\section{Results}

\subsection{The redshift distribution}

\begin{figure}
\begin{center}
\includegraphics[width=1.0\hsize,angle=0] {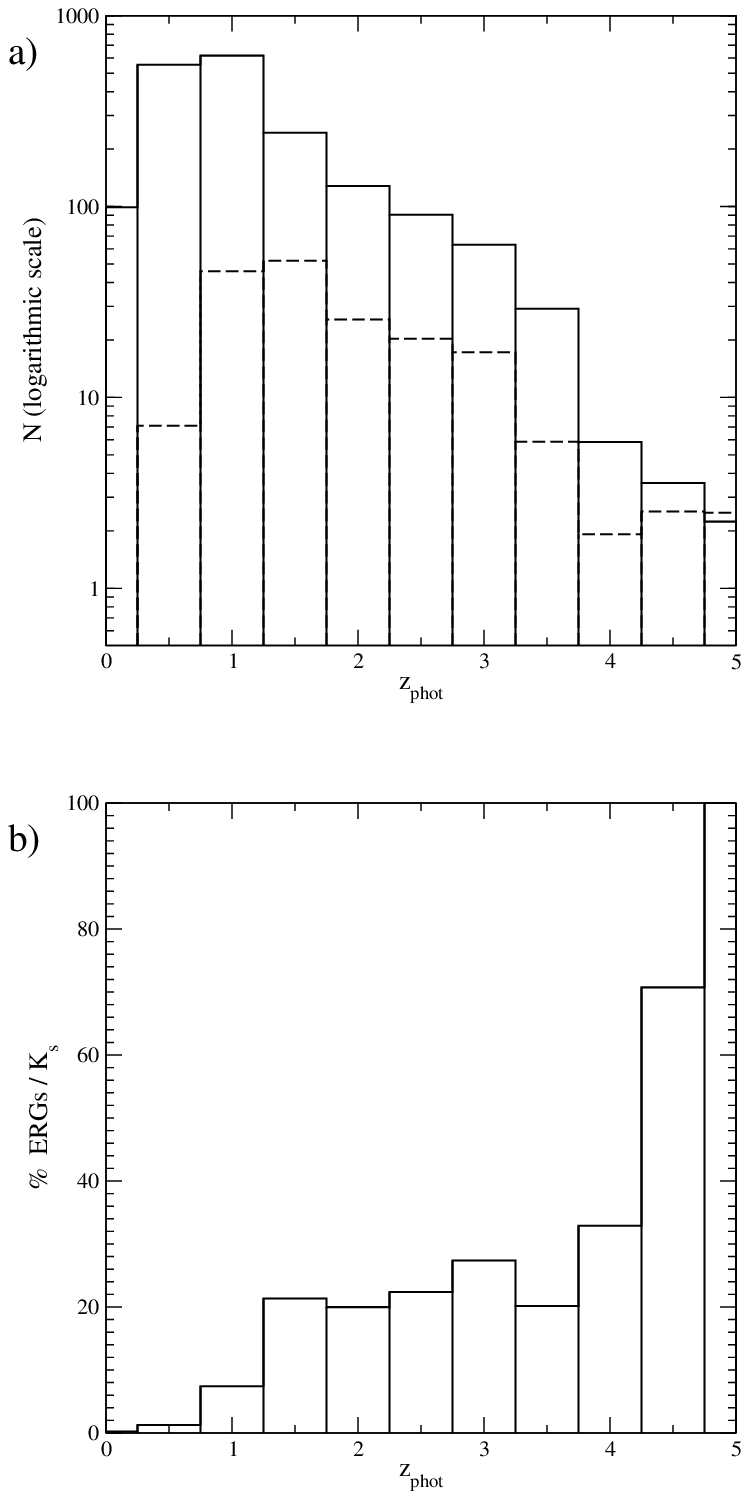}
\caption[]{\label{zdist} a) The redshift distribution of the total $\rm K_s \leq 22$ sample of galaxies (solid lines) compared with the redshift distribution of  the R03 ERGs (dashed lines). The adoption of a probability density distribution in redshift space for each source and incompleteness correction factors produce a non-integer number of sources in each redshift bin. b) The percentage of ERGs found among $\rm K_s$-selected galaxies as a function of redshift. }
\end{center}  
\end{figure}

   Figure \ref{zdist}(a) shows the redshift distribution of the total sample of $\rm K_s$-selected galaxies (solid lines) compared with the redshift distribution for the R03 ERGs obtained by C04 (dashed lines). The corresponding histograms have been constructed taking into account the probability density distribution in redshift space for each object, as given by HYPERZ in the output.  The consideration of probability densities yields more realistic representations of the redshift distributions, because they take into account all the uncertainties and degeneracies inherent to photometric redshift techniques. For the minority ($<$9\%) of objects for which we adopted the HYPERZ secondary solution or a BPZ/spectroscopic replacement, we only took into account a single redshift determination, as any information on a well-determined probability density distribution in redshift space was missing for these objects. Both histograms in Figure \ref{zdist}(a) include corrections for incompleteness, i.e. we multiplied by a factor 1.12, 1.16 and 1.25 the contribution of sources with $\rm 20.0 < K_s \leq 21.0$, $\rm 21.0 < K_s \leq 21.5$ and $\rm 21.5 < K_s \leq 22.0$, respectively. The distribution of the total $\rm K_s$-selected galaxy sample peaks at redshift $z_{phot}\sim 0.5-1.0$, while the ERG distribution has a maximum at  $z_{phot} \sim 1.5$. The difference in the redshift distributions is a consequence of the ERG colour cutoff. Figure \ref{zdist}(b) shows the relative fraction of ERGs among $\rm K_s$-selected galaxies, allowing for an easier comparison between both populations. We see that the fraction of $\rm K_s$-selected galaxies which are $\rm (I_{775}-K_s) > 3.92$ ERGs is approximately constant ($\sim$ 20\%-30\%) between redshifts $1.5 \lsim z_{phot} \lsim 4.0$. We only find two sources with estimated redshifts $z_{phot}>4$ in our present sample and these are the same two ERGs reported as confident $z>4$ candidates in C04.  

\subsection{The Hubble diagram}

\begin{figure}
\begin{center}
\includegraphics[width=1.0\hsize,angle=0] {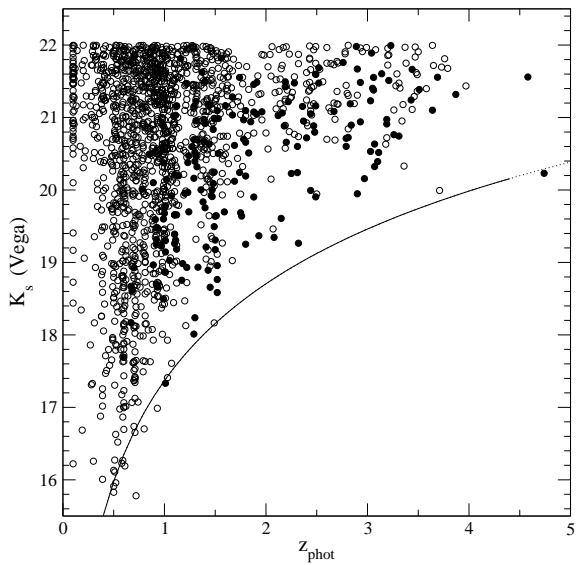}
\caption[]{\label{hubble} Total raw $\rm K_s$ magnitudes versus photometric redshifts $z_{phot}$. The empty circles correspond to the $\rm K_s$$\le$22-selected galaxies  which are not in the R03 ERG sample. The filled circles correspond to the ERGs in the R03 sample, with the redshifts estimated by C04. The solid line shows the empirical K-z relation for radio galaxies as obtained by Willott et al. (2003), which approximately coincides with the passive evolution of a $\rm 3L^\ast$ starburst formed at redshift $z=10$. The dotted line is a nominal extrapolation of the same law.}
\end{center}  
\end{figure}

  In Figure \ref{hubble} we plot the observed $\rm K_s$ magnitude versus photometric redshift $z_{phot}$ for the R03 ERGs (filled circles), as they were obtained by C04, and for all the other $\rm K_s$$\leq$22-selected galaxies (empty circles). The solid line shows the empirical  K-$z$ relation for radio galaxies  obtained by Willott et al. (2003), which approximately corresponds to the passive evolution of a  $\rm 3L^\ast$ starburst formed at redshift $z=10$ and indicates the behaviour of the most massive galaxies formed at very high redshifts. The redshift associated with each $\rm K_s$-selected galaxy in the present sample is either a HYPERZ solution or a BPZ/spectroscopic replacement, in accord with the criteria explained in Sections 3.1 and 3.2. 

  The comparison of the $\rm K_s$-$z_{phot}$ relation for extremely red and bluer $\rm K_s$-selected galaxies yields  two main conclusions. First, ERGs show the same large dispersion in the Hubble diagram as all the other $\rm K_s$-selected galaxies (although  the ERG colour cutoff produces the absence of these objects at the lower redshifts). Second, as was discussed in C04, the Hubble diagram is characterised by a lack of objects near the radio galaxy line at redshifts $z_{phot} \gsim 3$. We see that this effect is still present when all the other $\rm K_s$-selected galaxies are included.  However, we note the presence of two relatively bright objects at  $z_{phot}>3$ which are not $\rm (I_{775}-K_s)>3.92$ ERGs. These bright objects at $z_{phot}>3$ have $\rm (J-K_s)>2.2 \,(Vega)$, but some of them are missed when a typical ERG  $\rm (I_{775}-K_s)$ colour cut is applied (in the present case, in particular, these two sources have $\rm (I_{775}-K_s)<3.7$). These objects could correspond to galaxies in the final stages of a continuous but decaying process of star formation, sufficiently old ($\rm \sim 1-2 \, Gy$)  to have developed a $\rm 4000 \AA$-break which produces red observer-frame $\rm (J-K_s)$ colours. Yet, in some of them, the still rather active star formation produces considerable amounts of ultraviolet (UV) flux which, in conjunction with only modest dust extinction,  prevents these sources from displaying extremely red  $\rm (I_{775}-K_s)$ colours. In Section 4.4, we present some further discussion on the influence of the selection effects on the inclusion of the most massive galaxies when constructing $\rm K_s$-band galaxy samples.

\subsection{The evolution of the $\rm K_s$-band/ERG luminosity functions}

   In this section, we study the rest-frame $\rm K_s$-band LF for the ERGs and the total sample of $\rm K_s$-selected galaxies at different redshifts, in order to investigate any possible difference in their evolution. We extrapolated the rest-frame $\rm K_s$-band magnitudes using the best-fit SED of each galaxy. These extrapolations can be large at high redshifts, but the k-corrections in the $\rm K_s$-band are less dramatic than those corresponding to other bands (Poggianti 1997).  Although the computation of a rest-frame optical LF would require smaller extrapolations at high redshifts, a proper comparison with the ERG LF is difficult to perform at rest-frame optical wavelengths.  The values obtained for the ERG LF would strongly depend on the extinction corrections $A_V$ and, thus, for our purpose, a rest-frame optical LF would be more model-dependent than the LF in the rest-frame $\rm K_s$-band.
   
   We computed the $\rm K_s$-band LF binning the $\rm K_s \le 22$-sample of 1663 galaxies in both redshift and absolute magnitude space, in the same way as was done for the ERG sample by C04.  A single redshift estimate has been considered for each source (either the HYPERZ primary, the secondary, or a BPZ/spectroscopic solution, as detailed in Section 3.2). In the cases in which we adopted the HYPERZ primary solution, the corresponding k-corrected absolute magnitude $\rm M_{K_s}$ was directly obtained from the HYPERZ output. In all the other cases, where the estimated redshifts are relatively low, we computed the absolute magnitude $\rm M_{K_s}$ assuming that the k-corrections have negligible dependence on the SED shape (an assumption particularly valid for the $\rm K_s$ band at relatively low redshifts).  To take into account the limits of the survey ($\rm K_s=22$), we weighted each source by a factor $V_{maxbin}/V_{maxobs}$, where $V_{maxbin}$ is the volume determined by the  maximum redshift of the corresponding  bin and $V_{maxobs}$ is the volume determined by the maximum redshift at which the source would still be detected in the survey (provided it is lower than the maximum redshift of the bin). We also applied weighting factors of 1.12, 1.16 and 1.25 for sources with $\rm 20.0 < K_s \leq 21.0$, $\rm 21.0 < K_s \leq 21.5$ and $\rm 21.5 < K_s \leq 22.0$, respectively, to correct for incompleteness of the sample  (Section 2.2).

\begin{figure}
\begin{center}
\includegraphics[width=1.0\hsize,angle=0] {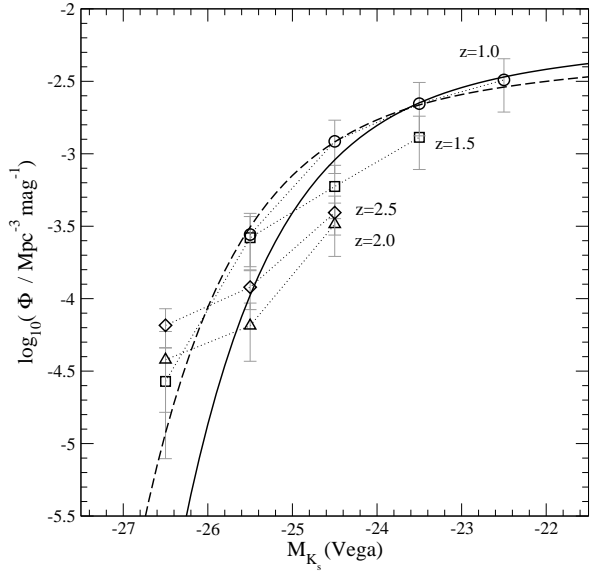}
\caption[]{\label{kslf} The rest-frame $\rm K_s$-band luminosity function at redshifts $\langle z_{phot}\rangle=1.0, 1.5, 2.0$ and $2.5$ (circles, squares, up triangles and diamonds, respectively). The values shown are strictly above the completeness limits of each redshift bin.  The solid line corresponds to the Schechter function fitted to the local K-band LF measured on 2MASS data by Kochanek et al. (2001). The dashed line shows the evolution of the  LF at redshift $\rm z=1$, as estimated by Drory et al. (2003).}
\end{center}  
\end{figure}

\begin{figure}
\begin{center}
\includegraphics[width=1.0\hsize,angle=0] {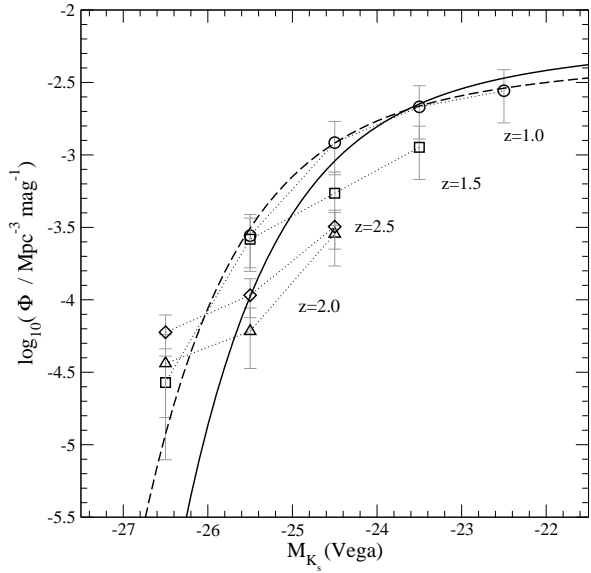}
\caption[]{\label{rawkslf} The raw rest-frame $\rm K_s$-band luminosity function at redshifts $\langle z_{phot}\rangle=1.0, 1.5, 2.0$ and $2.5$ (no $V_{max}$ corrections, no corrections for incompleteness). The symbols are the same as in Figure \ref{kslf}.}
\end{center}  
\end{figure}

\begin{figure}
\begin{center}
\includegraphics[width=0.8\hsize,angle=0] {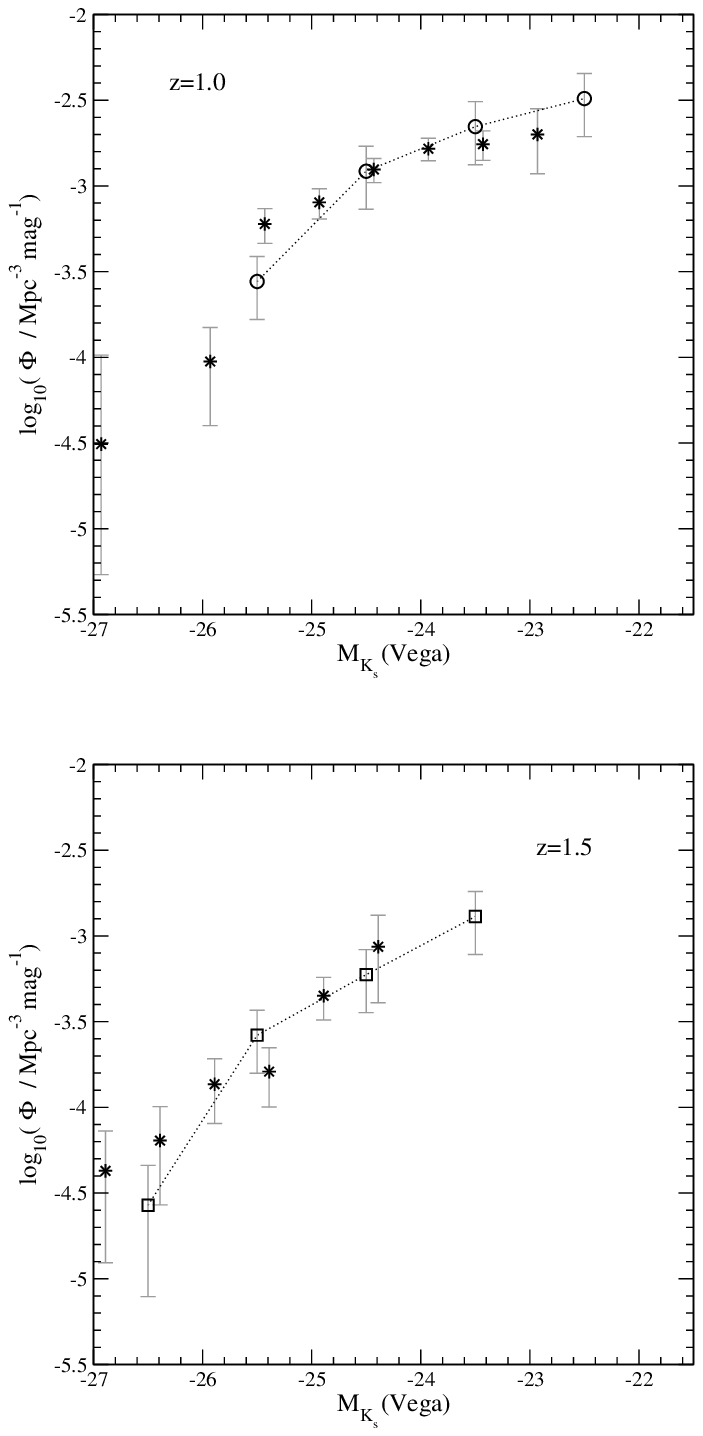}
\caption[]{\label{k20lfcomp}  The comparison of the rest-frame $\rm K_s$-band LF obtained from the present study (empty symbols) with the LF obtained by Pozzetti et al. (2003) using K20 survey data (asterisks): a) $\langle z_{phot}\rangle=1.0$; b) $\langle z_{phot}\rangle= 1.5$.}
\end{center}  
\end{figure}

\newpage

\begin{figure*}
\begin{center}
\includegraphics[width=1.0\hsize,angle=0] {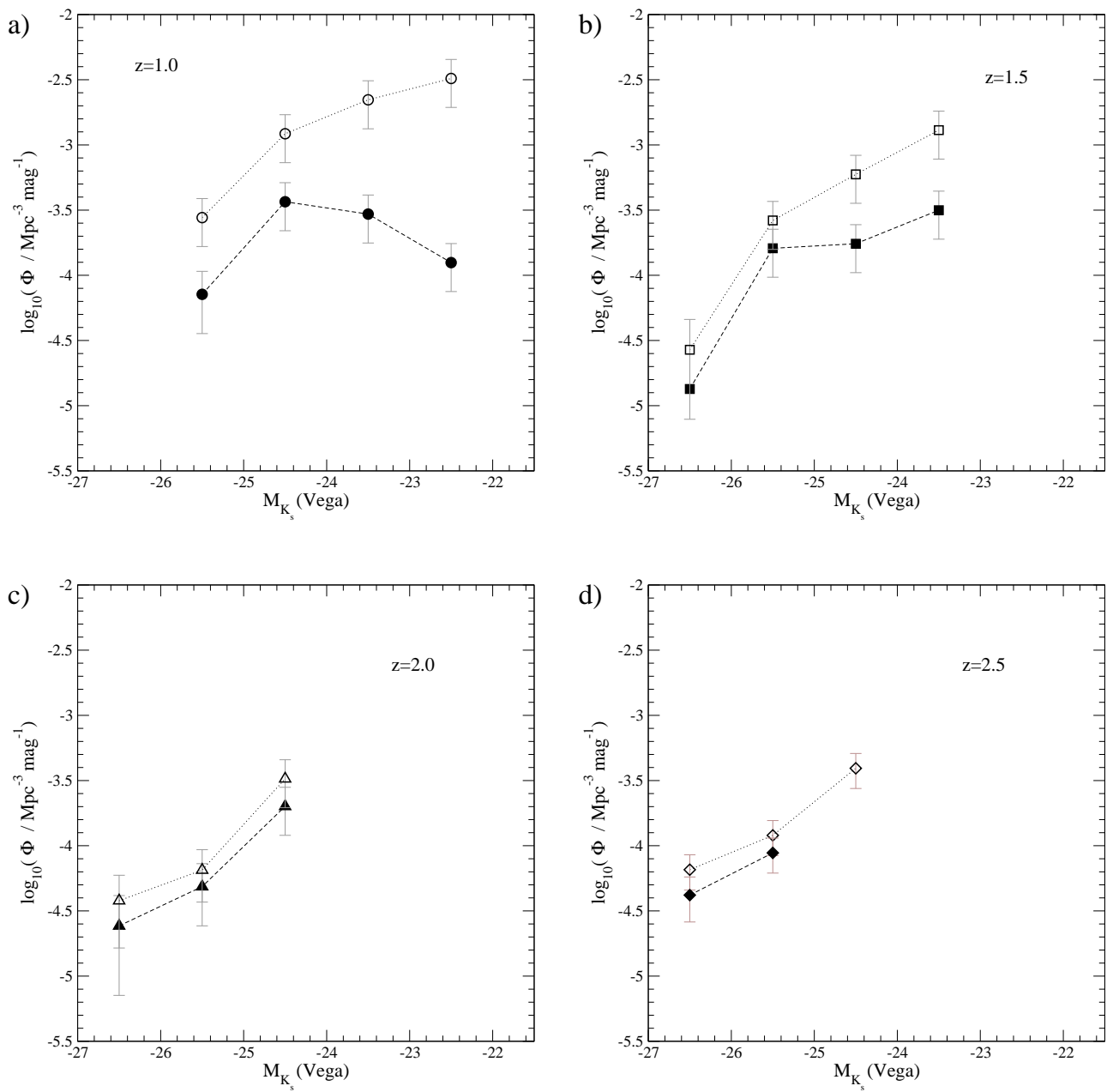}
\caption[]{\label{ksero_lf} Compared evolution of the total $\rm K_s$-band (empty symbols) and the ERG (filled symbols) LFs with redshift. The values shown are strictly above the completeness limits of each redshift bin. The error bars for each point are the maximum of Poissonian errors and the errors due to cosmic variance.} 
\end{center}  
\end{figure*}

\begin{figure*}
\begin{center}
\includegraphics[width=1.0\hsize,angle=0] {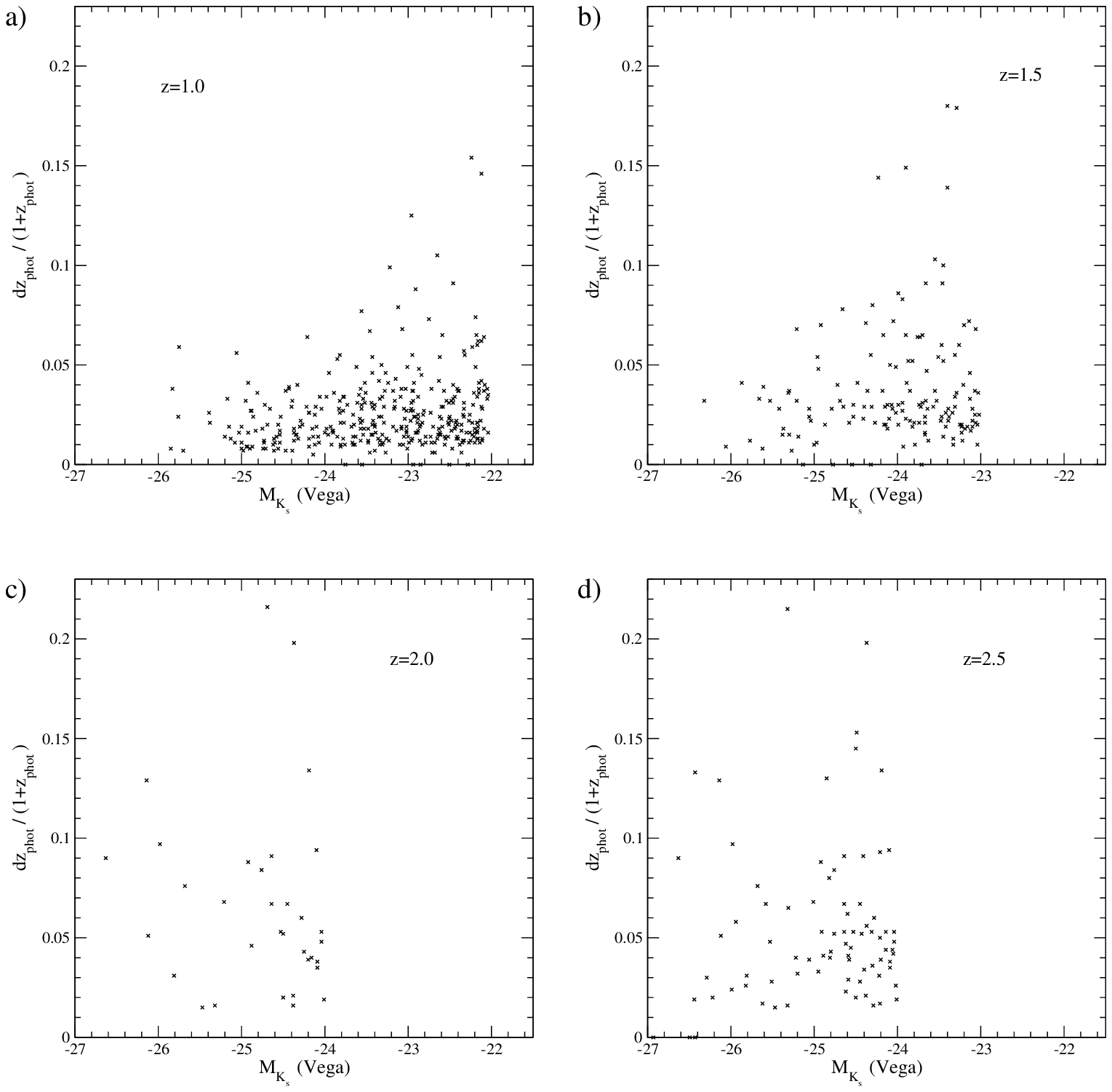}
\caption[]{\label{errlf} The relative errors of the redshift estimates for the sources considered in the computation of the $\rm K_s$-band LF at different redshifts. The redshift bins considered in a), b) and c) have width $\Delta z_{phot}=0.5$, while the redshift bin considered in d) has width $\Delta z_{phot}=1$.} 
\end{center}  
\end{figure*}

  Figure \ref{kslf} shows the total $\rm K_s$-band LF  at redshifts $\langle z_{phot}\rangle=1.0, 1.5, 2.0$ and $2.5$ (circles, squares, up triangles and diamonds, respectively). The values shown are strictly above the completeness limits of each redshift bin. The error bars correspond to the maximum of Poissonian errors and the errors due to cosmic variance, which we considered on average as 40\% and 30\% of the number counts at redshifts $z<2$ and $2<z<3$, respectively (Somerville et al. 2004). Although not all the $\rm K_s$-selected galaxies are expected to display the strong clustering observed in ERGs (R03), we adopted the same cosmic variance error bars as in C04 for conservative reasons. We do not present the LF at redshift $\langle z_{phot}\rangle=0.5$ because our surveyed volume  is small at those redshifts and the LF is quite affected by the presence of large scale structure at $z=0.67-0.73$ (Le F\`evre et al. 2004; Vanzella et al. 2004). We show, for comparison, the local K-band LF obtained by Kochanek et al. (2001) using  Two Micron All Sky Survey (2MASS) data, and its evolution to redshift $z=1$ as estimated by Drory et al. (2003) (solid and dashed lines, respectively). Drory et al. (2003) found that the density of bright objects increases from redshift $\rm z=0$ to $\rm z=1$. A similar trend was found by Pozzetti et al. (2003) from the analysis of the near-IR LF up to redshift $\langle z\rangle=1.5$, obtained using K20 survey data. Our results confirm this, and show  no evidence for the bright end to decrease again up to at least redshifts  $\langle z_{phot}\rangle=2.5$. The same effect was observed for the ERG LF by C04. Figure \ref{kslf} confirms that the behaviour of a constant (or even possibly increasing) evolution for the bright end of the LF is a property of the total $\rm K_s$-selected galaxy population. 
 At this point, we should remind the potential effects of the lack of deep U-band data on our estimated redshifts. As we discussed in Section 3.2, a non-negligible fraction of the high-redshift sample analysed here could  potentially be contaminated with lower redshift sources with degenerate solutions in photometric redshift space. This fact could partially bias the trend we observe in the bright end of the LFs.    At intermediate magnitudes, on the other hand, our results suggest a non-negligible decrease in the density of objects, although wider and deeper surveys are necessary to better quantify this evolution. To demonstrate that none of our conclusions depend on the $V_{max}$ corrections or the corrections for incompleteness, we show in Figure \ref{rawkslf} the `raw' $\rm K_s$-band LF, i.e. the LF constructed without applying any of these corrections. We note that the analysis of the raw LF allows us to extract exactly the same conclusions as in the case of the corrected version presented in Figure  \ref{kslf}. The comparison of our rest-frame $\rm K_s$-band LF with that obtained for the K20 survey is shown in Figure \ref{k20lfcomp}.

   The four panels in Figure \ref{ksero_lf} compare the evolution of the total (empty symbols) and the ERG (filled symbols) $\rm K_s$-band LFs with redshift.  The latter is the same as presented in C04, although here we only show our LF values strictly above the completeness limits of each redshift bin. For each LF individually, we adopted the maximum of the Poisson errors and the errors due to cosmic variance.  Any possible error produced by uncertainties in our redshift estimations is expected to be contained within the Poisson/cosmic variance error bars. Figure \ref{errlf} shows the relative error $dz_{phot}/(1+z_{phot})$ in the redshift estimation of each one of the sources making the $\rm K_s$-band LF at different redshifts. The absolute error $dz_{phot}$ corresponds to 1$\sigma$-confidence levels, as obtained from HYPERZ. The sources with $dz_{phot}=0$ correspond to cases in which the HYPERZ secondary solution has been adopted for the redshift estimate.  In every redshift bin, the median of the relative errors for the redshift estimates is $dz_{phot}/(1+z_{phot}) < 0.06$ and the number of sources with $dz_{phot}/(1+z_{phot})>0.2$ is very small ($<$3\% in all cases).

   A clear feature in the evolution of the LF is the increasing role of ERGs in reproducing the bright end of the total $\rm K_s$-band LF with increasing redshift, reaching a maximum at $\langle z_{phot}\rangle=2.0$, where a considerable fraction of the brightest $\rm K_s$-selected galaxies are ERGs. Also, from the comparison of the total and the ERG $\rm K_s$-band LFs at redshift $\langle z_{phot}\rangle=1.0$,  we conclude that the latter has a bell-like shape, consistent with a Schechter function with a flat slope. This result is in agreement with the different shapes determined for the LFs of red and blue galaxies at different redshifts (e.g. Lilly et al. 1995; Pozzetti et al 2003; Kodama et al. 2004). At redshift  $\langle z_{phot}\rangle=1.5$, the total $\rm K_s$-band and the ERG LFs are more similar, but a significant density of blue galaxies is still necessary  to account for the  total $\rm K_s$-band galaxy population at intermediate magnitudes. At $2 \lsim z_{phot} \lsim 3$, our completeness limits only allow us to explore the bright end of the $\rm K_s$-band LF, which appears mainly dominated by the ERGs.

\subsection{The evolution of massive galaxies}

     In C04, it was shown that the comoving densities of ERG progenitors of the local $\rm L>L^\ast$ galaxy population, under passive evolution, were below the values expected by the evolution of dark matter haloes in a  $\Lambda$-Cold Dark Matter ($\Lambda$CDM) formalism (Kauffmann \& Charlot 1998; Somerville \& Primack 1999). This fact has been interpreted as either a deficiency of the ERG population to account for all the progenitors of the local  $\rm L>L^\ast$ galaxies, or as due to the incomplete picture described by a passive evolution scenario. Our aim here is to further test these two possibilities using a similar approach: studying the evolution of the comoving number density of massive systems and, in general, the evolution of the stellar mass content of the Universe with redshift. The evolution of the stellar mass density has recently been studied in different fields up to redshift $z \sim 2$  (Glazebrook et al. 2004, Drory et al. 2004, Fontana et al. 2004), and in small-area surveys up to redshift $z \sim 3$ (Dickinson et al. 2003, Fontana et al. 2003, Rudnick et al. 2003). Here we extend the analysis of the comoving mass density evolution up to redshift $z \sim 4$, as obtained from the study of our deep sample of $\rm K_s$-selected galaxies in the GOODS/CDFS field.

       We computed the rest-frame $\rm K_s$-band luminosity  of each of our galaxies using the k-corrected absolute magnitude $\rm M_{K_s}$, associated with the HYPERZ primary or secondary solution, on a case-by-case basis. In the cases in which we adopted the HYPERZ secondary solution, the absolute magnitude $\rm M_{K_s}$ has been extrapolated as we explained in Section 4.3.   To estimate the mass of our galaxies, we used, in each case, the galaxy luminosity and a mass-to-light ($\rm M/L_{K_s}$) ratio depending on the corresponding best-fit HYPERZ SED type and age. For the eight objects with BPZ/spectroscopic redshifts, any information on the galaxy SED type or age is missing, so  these sources are not taken into account in all the analysis made in this section.  We used the public code GALAXEV (Bruzual \& Charlot 2003) to construct  a grid of mass-to-light ($\rm M/L_{K_s}$) ratios depending on galaxy age. In the rest-frame $\rm K_s$-band, these ratios have little dependence on either the dust corrections or the galaxy star formation history. Thus, to model the $\rm M/L_{K_s}$ ratios as a function of age, we used only two galaxy templates, sufficiently representative of the different HYPERZ library SEDs. Both GALAXEV templates had a solar metallicity and no dust corrections, but corresponded to two different exponentially-declining star formation histories: fast ($\rm \tau \sim 0.1 \, Gy$) and slow ($\rm \tau \sim 5 \, Gy$).  The final $\rm M/L_{K_s}$ value adopted for each galaxy was the one corresponding to the GALAXEV template most similar to the best-fit HYPERZ SED, and to the best-fit age, obtained by interpolating between the $\rm M/L_{K_s}$ values in our age grid. The sufficient sampling of our age grid makes that the  interpolation in age does not constitute a significant source of uncertainty. The interpolation in star formation history, on the contrary,  might constitute a non-negligible source of uncertainty on the estimated masses. However,  for the two adopted   star formation histories, the $\rm M/L_{K_s}$ ratios only differ by a factor $<2$ at any given age. These differences are considerably smaller than those obtained with optical $\rm M/L$ ratios, even with the use of more refined grids. On the other hand, it should be noted that the uncertainities introduced in our mass estimates by the use of extrapolated luminosities in the rest-frame $\rm K_s$-band are not larger than the uncertainties produced linking mass and light in rest-frame optical bands: in both cases, the precision in the estimated mass for each galaxy relies on the precision of its SED modelling. We adopted a Salpeter initial mass function (IMF), for ease of comparison with most of the values quoted in the literature. To compute comoving densities, we used the complete probability distribution in redshift space to determine the potential contribution of each object at different redshifts (note, however, that a more rigourous procedure should take into account a probability density  in the whole parameter space). For the sources for which we adopted the HYPERZ secondary solution, a single redshift has been  considered, as in these cases we cannot recover a well-determined probability density function.

\begin{figure}
\begin{center}
\includegraphics[width=1.0\hsize,angle=0] {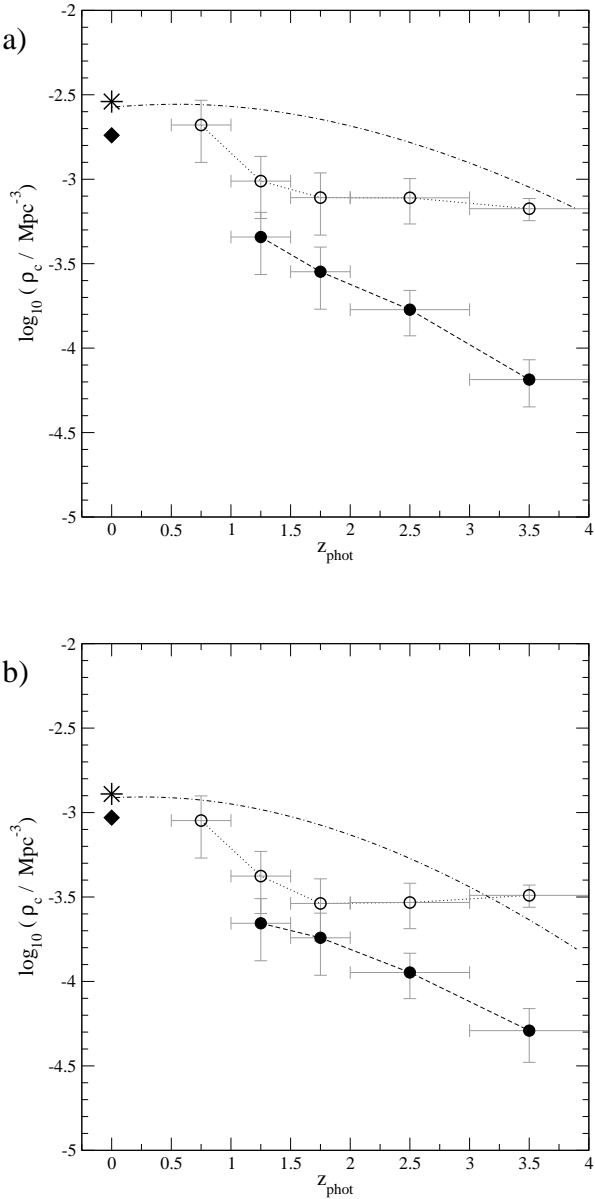}
\caption[]{\label{cnumd} The comoving number density of galaxies with stellar mass $\rm M>5 \times 10^{10} M_\odot$ (a) and $\rm M>1 \times 10^{11} M_\odot$ (b). The empty circles indicate the densities of all the $\rm K_s$-band selected galaxies, while the filled circles show the contribution of the ERGs selected in the same field. In both plots (a) and (b), the star-like symbol and diamond indicate the corresponding local values for all and early-type galaxies, respectively (Bell et al. 2003).  The dotted-dashed lines indicate the evolution of the number density of haloes with total mass $\rm M>8 \times 10^{11} M_\odot$ (a) and $\rm M>2 \times 10^{12} M_\odot$ (b), as obtained from $\Lambda$CDM models of structure formation. }
\end{center}  
\end{figure}

      Figure \ref{cnumd} shows the comoving number densities of galaxies which have assembled a stellar mass $\rm M>5 \times 10^{10} M_\odot$ (a) and $\rm M>1 \times 10^{11} M_\odot$ (b), as a function of redshift. The empty circles correspond to all the $\rm K_s$-band selected galaxies, while the filled circles indicate the contribution of the R03 ERGs selected in the same field. The horizontal error bars indicate the binning in redshift space. The vertical error bars correspond to cosmic variance (40\%, 30\% and 15\% of the number counts at redshifts $z_{phot}<2$, $2<z_{phot}<3$ and $3<z_{phot}<4$, respectively), which is the dominant source of error in this case.   $V_{maxbin}/V_{maxobs}$ and incompleteness correction factors have been applied throughout. In both panels (a) and (b), the star-like symbol (diamond) indicates the local number densities of all (early-type) galaxies with stellar masses above the corresponding mass cuts. The local number densities have been obtained by integrating the K-band-derived mass functions computed by Bell et al. (2003), converted to a Salpeter IMF.
      
      From inspection of  Figure \ref{cnumd}, we see that a considerable fraction ($\sim$ 20\%-25\%) of the galaxies with mass $\rm M>5 \times 10^{10} M_\odot$ and $\rm M>1 \times 10^{11} M_\odot$ is in place before redshift $z_{phot} \sim 4$. The number density is virtually constant down to redshift $z_{phot} \approx 2$ and  the assembly of the remaining massive systems appears to have been produced at redshifts  $z_{phot} \lsim 1.5$. It should be noted that the fraction of massive systems estimated to be at $z \gsim 2$ would be somewhat overestimated if  lower redshift contaminants were present in the high-redshift sample. Thus, the percentages quoted here should be regarded as upper limits on the fraction of massive systems present at high redshifts. The evolution observed in the number densities of sources with mass $\rm M>5 \times 10^{10} M_\odot$ and $\rm M>1 \times 10^{11} M_\odot$ suggests the existence of a two-fold  mechanism for the construction of massive galaxies: while most of the systems are assembled at relatively low redshifts, as expected in hierarchical models of galaxy formation, a  substantial fraction is assembled very efficiently at very early times.

       As in C04, we compare the comoving number densities of  massive galaxies with the corresponding densities of dark matter haloes massive enough to host these systems, as they are predicted by $\Lambda$CDM models of structure formation. In Figure \ref{cnumd}, the dotted-dashed lines represent the expected evolution of the number densities of haloes with total masses $\rm M>8 \times 10^{11} M_\odot$ (a) and $\rm M>2 \times 10^{12} M_\odot$ (b).  The halo mass thresholds have been deliberately selected to coincide with the local number densities of galaxies with stellar mass $\rm M>5 \times 10^{10} M_\odot$ and $\rm M>1 \times 10^{11} M_\odot$, respectively. In the case of a single halo occupation number, we would expect the evolution of the number densities of galaxies to follow the evolution of the corresponding number density of haloes. This is what we observe up to redshift $z_{phot} \sim 1$. However, between redshifts $z_{phot} \sim 1$ and $z_{phot} \sim 3$, the predicted number densities of host haloes are significantly larger than the observed densities of already assembled massive galaxies. The latter are approximately constant up to redshift $z_{phot} \sim 4$, while the number of haloes decrease with redshift, in such a way that the number of galaxies and haloes coincide again at $z_{phot} \gsim 3$. This comparison suggests that, at very high redshift,  the timescale in which halo merging and galaxy collapse occur are quite similar and, thus, the number density of galaxies traces the number density of haloes. At  $z_{phot} \lsim 3$, on the contrary, halo merging could be faster than galaxy collapse, explaining why the observed density of massive galaxies is only a fraction of the density of the corresponding haloes. By  $z_{phot} \lsim 1$, the assembly of massive galaxies is mostly complete, and the number of galaxies and haloes coincides again.  We note that, even while we have an excess of galaxies with mass  $\rm M>1 \times 10^{11} M_\odot$ with respect to the corresponding number of haloes at  $z_{phot} \gsim 3$, this excess is not significant and could be contained within the error bars if the error of 15\% in the number counts were somewhat underestimating the cosmic variance at those redshifts. Future deeper surveys will be able to test whether the predicted number densities of haloes is still compatible with the observational data beyond redshift  $z_{phot} \sim 4$, and set tighter constraints on the high formation redshifts of the oldest massive systems.

       As we stated at the beginning of this section, one of our main aims is to assess the role of ERGs to account for the progenitors of the most massive systems and the validity of the passive evolution assumption. We make clear that, of course, only a subset of ERGs can potentially be the progenitors of the most massive local galaxies. The whole of the ERG population spans a wide variety of redshifts and masses (C04), and many ERGs are completely irrelevant for the present discussion.  The comparison of the number densities for the massive ERGs and all the massive $\rm K_s$-selected galaxies in Figure \ref{cnumd}(a) and \ref{cnumd}(b) illustrates  the fact that ERGs trace the high mass end of the  $\rm K_s$-selected galaxies. However, we observe that ERGs cannot account for all the systems with mass $\rm M>1 \times 10^{11} M_\odot$.  At redshifts $z_{phot} \sim 1-2$, ERGs constitute $\sim$ 50\%-70\% of all the $\rm K_s$-selected galaxies which have assembled a stellar mass  $\rm M>1 \times 10^{11} M_\odot$.  Figure \ref{cnumd}(b) also suggests the existence of an evolutionary sequence between massive $\rm K_s$-selected galaxies at very high redshift and massive ERGs. We note that the comoving number densities of $\rm K_s$-selected galaxies at $\langle z_{phot} \rangle$=3.5 is only slightly larger than the number density of  ERGs with similar mass at redshift $z_{phot} \sim 1$, indicating that most of the massive systems at very high redshift might have evolved into ERGs by redshift $z_{phot} \sim 1$. The extremely red colours could be the consequence of the ageing of the stellar populations and this would be evidence for passive evolution since very high redshift.  However, as stated in C04, the colours of some of the reddest ERGs can only be explained by the superposition of an evolved stellar population and dust. This would indicate that, in some of these galaxies, the passive ageing of the stellar populations could be interrupted by the production of new star formation. But, except for these possible temporary periods of additional star formation, ERGs seem to account for the fraction of massive galaxies at high redshift which have plausibly been evolving  under passive evolution. In any case, we see that the massive galaxies which  could have evolved passively since very high redshifts are only a fraction of the local value and, thus, a passive evolution scenario alone is not able to explain the construction of all the massive systems.
      
      At $z_{phot}>2$, on the contrary, we find a significant fraction of $\rm K_s$-selected galaxies with mass  $\rm M>1 \times 10^{11} M_\odot$  which are not $\rm (I_{775}-K_s)$-selected ERGs.  Regarding the effects of completeness, we estimate that our $\rm K_s$-selected sample is complete for galaxies with mass $\rm M>1 \times 10^{11} M_\odot$ up to redshift $z_{phot} \sim 4$, based on the median of the k-corrections and the maximum possible $\rm M/L_{K_s}$ ratio at that redshift. The median of the k-corrections is somewhat larger when only the ERGs are considered, and we estimate that the ERG sample is complete to a mass $\rm M=1 \times 10^{11} M_\odot$ up to redshift $z_{phot} \sim 3$. Thus, incompleteness only affects the comparison of the ERG and all the $\rm K_s$-selected galaxy densities in the highest redshift bin.  Most of our massive galaxies at $z_{phot}>2$ have $\rm (J-K_s)>2$,  indicating that a red $\rm (J-K_s)$ colour cut provides a more efficient method to select massive systems at very high redshift than an $\rm (I_{775}-K_s)$ cut (Franx et al. 2003). In Section 5, we predict the IR colours of our $\rm K_s$-selected sources in different Spitzer/IRAC channels and present some further discussion on the efficiency of different pure IR colours to select high redshift galaxies.

\begin{figure}
\begin{center}
\includegraphics[width=1.0\hsize,angle=0] {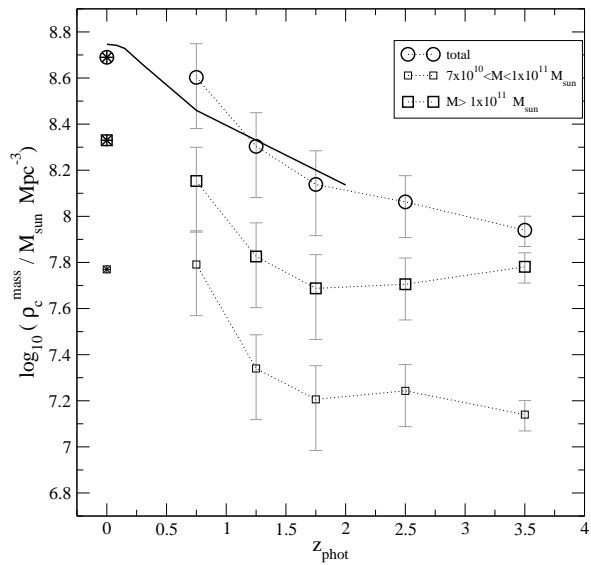}
\caption[]{\label{massbudg}  The distribution of the stellar mass density as a function of redshift, as derived from our deep sample of $\rm K_s$-selected galaxies. The circles indicate the  total stellar mass density. The large (small) squares are the contributions of galaxies with masses $\rm M>1 \times 10^{11} M_\odot$ ($\rm 7 \times 10^{10} M_\odot< M <1 \times 10^{11} M_\odot$), for which our $\rm K_s=22$ -limited sample is estimated to be complete up to redshift $z_{phot} \approx 4$ ($z_{phot} \approx 3$). The symbols with a star at $z_{phot}=0$ show the corresponding local values obtained from the integration of the local galaxy stellar mass function (Cole et al. 2001, Bell et al. 2003). The solid line is the evolution of the stellar mass density as obtained from the integration of  the star-formation rates derived from Sloan Digital Sky Survey (SDSS) data (Heavens et al. 2004). }
\end{center}  
\end{figure}

\begin{figure}
\begin{center}
\includegraphics[width=1.0\hsize,angle=0] {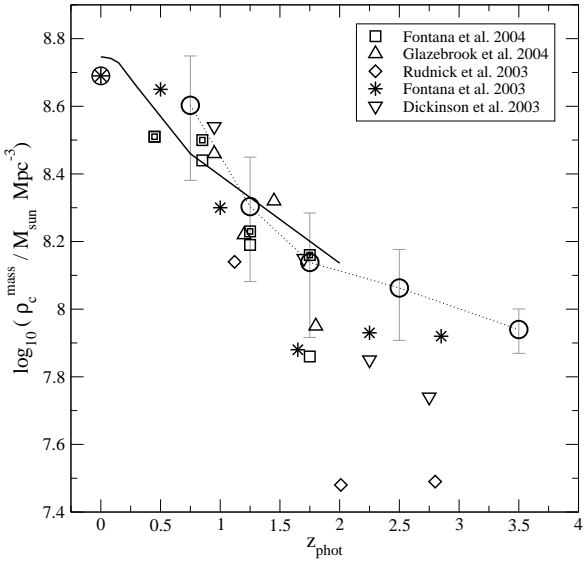}
\caption[]{\label{compstm}  The comparison of the total stellar mass densities obtained from the present study (empty circles, with error bars) with values previously obtained by other authors: Dickinson et al. 2003 (down-triangles); Fontana et al. 2003 (asterisks); Rudnick et al. 2003 (diamonds); Glazebrook et al. 2004 (up-triangles); Fontana et al. 2004 (squares and double squares for the observed and extrapolated values in the K20 survey, respectively).  The circle with a star at $z_{phot}=0$ show the local value of the total stellar mass density obtained from the integration of the local galaxy stellar mass function (Cole et al. 2001, Bell et al. 2003). The solid line is the evolution of the stellar mass density as obtained from the integration of  the star-formation rates derived from Sloan Digital Sky Survey (SDSS) data (Heavens et al. 2004). Dickinson et al. values correspond to the average of the values obtained with the 1-component and 2-component models with solar metallicity (see Table 3 in Dickinson et al. 2003). Rudnick et al. values are computed only on galaxies with rest-frame V-band luminosity $\rm L_V > 1.4 \times 10^{10} L_{\odot}$ and, thus, miss a significant fraction of the mass. Glazebrook et al. values only correspond to galaxies with estimated mass $\rm \log(M)>10.46$ and are at least partially incomplete above redshift $z \sim 1.2$. It should be noted that all the previous determinations of stellar mass densities at redshifts $z>2$ have been made on surveys with coverage areas $\sim 10$ times smaller than the area analysed in this study. All the values in this figure assume a Salpeter IMF and a cosmology with $\rm H_o=70 \,\rm km \, s^{-1} Mpc^{-1}$, $\rm \Omega_M=0.3$ and $\rm \Omega_\Lambda=0.7$.}

\end{center}  
\end{figure}

     Another point of interest is the analysis of how the total stellar mass budget is distributed at different redshifts. Figure \ref{massbudg} shows the evolution of the  total stellar mass density (circles) with  redshift, as derived from our sample of $\rm K_s$-selected galaxies. The large and small squares indicate the contribution of galaxies with massses $\rm M>1 \times 10^{11} M_\odot$ and $\rm 7 \times 10^{10} M_\odot< M <1 \times 10^{11} M_\odot$, respectively. (Our $\rm K_s=$22-limited sample is estimated to be complete for galaxies with masses $\rm M>1 \times 10^{11} M_\odot$ and $\rm M>7 \times 10^{10} M_\odot$ up to redshifts $z_{phot} \approx 4$ and $z_{phot} \approx 3$, respectively). The symbols with a star at $z=0$ show the  corresponding local values obtained from the integration of the local galaxy stellar mass function (Cole et al. 2001, Bell et al. 2003). The solid line is the evolution of the stellar mass density as expected from the integration of  the star-formation rates derived from Sloan Digital Sky Survey (SDSS) data (Heavens et al. 2004). From Figure \ref{massbudg}, we see that most of the stellar mass is assembled at redshifts $0.5 \lsim z_{phot} \lsim 1.5$, in agreement with the tendency previously observed by other authors. At $\langle z_{phot} \rangle= 1.75$, we find that the total stellar mass density is already $\sim$ 25\% - 30\% of the local value, a percentage somewhat higher than the value of $\sim$ 13\% obtained by Rudnick et al.(2003) at $z \sim 2$, using a small-area survey in the Hubble Deep Field South (HDFS), and more similar to the value of $\sim$ 20\% to $\sim$ 35\% determined by Fontana et al. (2004) from the K20 survey.  Our observed value of the total  stellar mass density at $\langle z_{phot} \rangle= 1.75$ is $\rm (1.37 \pm 0.55) \times 10^8 M_\odot Mpc^{-3}$ (taking into account cosmic variance), which is in excellent agreement with the value $\rm (1.45^{+0.41}_{-0.62})\times 10^8 M_\odot Mpc^{-3}$, as predicted by Fontana et al. (2004) by the extrapolation of the Schechter fits to the mass function of the K20 survey at these redshifts. It is worthwhile to note, however, that the observed mass density at $\langle z_{phot} \rangle= 1.75$ in a survey limited at $\rm K_s=20$ is only $\sim$ 50\% of the total value (Fontana et al. 2004), while a survey extended up to $\rm K_s=22$ is able to recover most of the mass at that redshift, as we see from the comparison with the assembled stellar mass obtained from the integration of the star-formation rates derived from SDSS data. Beyond $z_{phot} \sim 2$, we observe only a slow decrease of the total stellar mass density, in spite of the increasing incompleteness for the intermediate and lower mass galaxies. We  estimate that $\sim$ 24\% and $\sim$ 18\% of the local stellar mass was assembled at redshifts $\langle z_{phot} \rangle=2.5$ and $\langle z_{phot} \rangle=3.5$, respectively. A detailed comparison of our total stellar mass densities at different redshifts with other values found in the literature is presented in Figure \ref{compstm} (see figure caption for further references). It should be emphasized that all the determinations of stellar mass densities at redhsifts $z>2$ previous to this study have been made on pencil-beam surveys, with coverage areas $\rm \lsim 5 \, arcmin^2$.

     The analysis of the distribution of the stellar mass budget in Figure \ref{massbudg} also allows us to extract an interesting conclusion: a minimum of $\sim$ 45\% and a maximum of $\sim$ 70\% of the stellar mass at redshifts $z_{phot} \gsim 3$ is contained in galaxies with assembled mass $\rm M>1 \times 10^{11} M_\odot$. To compute the lower limit ($\sim$ 45\%), we considered the fraction of the stellar mass density assembled in galaxies with $\rm M>1 \times 10^{11} M_\odot$  at $\langle z_{phot}\rangle = 3.5$ over the total stellar mass density at $\langle z_{phot}\rangle = 2.5$. The latter can be considered as an upper limit to the total stellar mass density assembled at $\langle z_{phot}\rangle = 3.5$. The upper limit to the fraction of mass locked in massive systems ($\sim$ 70\%) has been obtained by directly comparing the stellar mass density corresponding to galaxies with $\rm M>1 \times 10^{11} M_\odot$ and the total value, both at $\langle z_{phot}\rangle = 3.5$.  The large percentage of stellar mass locked in massive systems is related to the significant number density of massive systems we find to be present before redshift $z_{phot} \sim 4$, and  indicates that galaxy/star formation at very high redshift is an extremely efficient process. These massive galaxies present at very  high redshift have presumably been formed from the early collapse of the high peaks of the density fluctuation field (e.g. Mo, Jing \& White 1997). Such regions of high density are predicted to be strongly clustered, a fact which is perfectly consistent with the strong clustering observed in red $\rm K_s$-selected galaxies at $2<z_{phot}<4$ (Daddi et al. 2003) and in the ERGs up to the faintest magnitudes (R03). Evidence for a similar `anti-hierarchical' behaviour in favour of biased galaxy formation has also been found by other authors from the study of galaxies at redshift $z \sim 1$ (Kodama et al. 2004).  Our results are consistent with a scenario, already suggested by previous studies, in which a significant fraction of the massive galaxies we see today have been assembled at very high redshifts ($z \gsim 4$), evolved to $\rm (I-K_s)$ ERGs by redshift $z \sim 1$ and possibly became members of massive galaxy clusters in the local Universe. This first detailed study of a significant deep sample of $\rm K_s$-selected galaxies allows us to show and summarise in one plot (Figure \ref{cnumd}b) the evolutionary line traced by the massive $\rm K_s$-selected galaxies and the massive ERGs since very high redshifts.


\section{Predictions of Spitzer/IRAC magnitudes}

   HYPERZ provides the possibility of extrapolating the galaxy magnitudes beyond the observed bands using the best-fit SED template for each object. We used the observer-frame best-fit templates of each of our $\rm K_s$-selected  galaxies to predict their expected magnitudes in three of the Spitzer/IRAC channels, i.e. $3.6 \mu m$, $4.5 \mu m$ and $5.8 \mu m$, as well as the derived IR colour distribution as a function of photometric redshifts.  We present in this section the predicted colour distributions for all our $\rm K_s$-selected galaxies. The complete catalog of $\rm K_s$-selected galaxies with the individual Spitzer/IRAC predicted magnitudes will be released in a future paper.   To convert from fluxes to magnitudes, we used the IRAC zero-flux points: 277.5Jy, 179.5Jy and 116.6Jy at $3.6 \mu m$, $4.5 \mu m$ and $5.8 \mu m$, respectively (Fazio et al. 2004). 
   
\begin{figure}
\begin{center}
\includegraphics[width=0.88\hsize,angle=0] {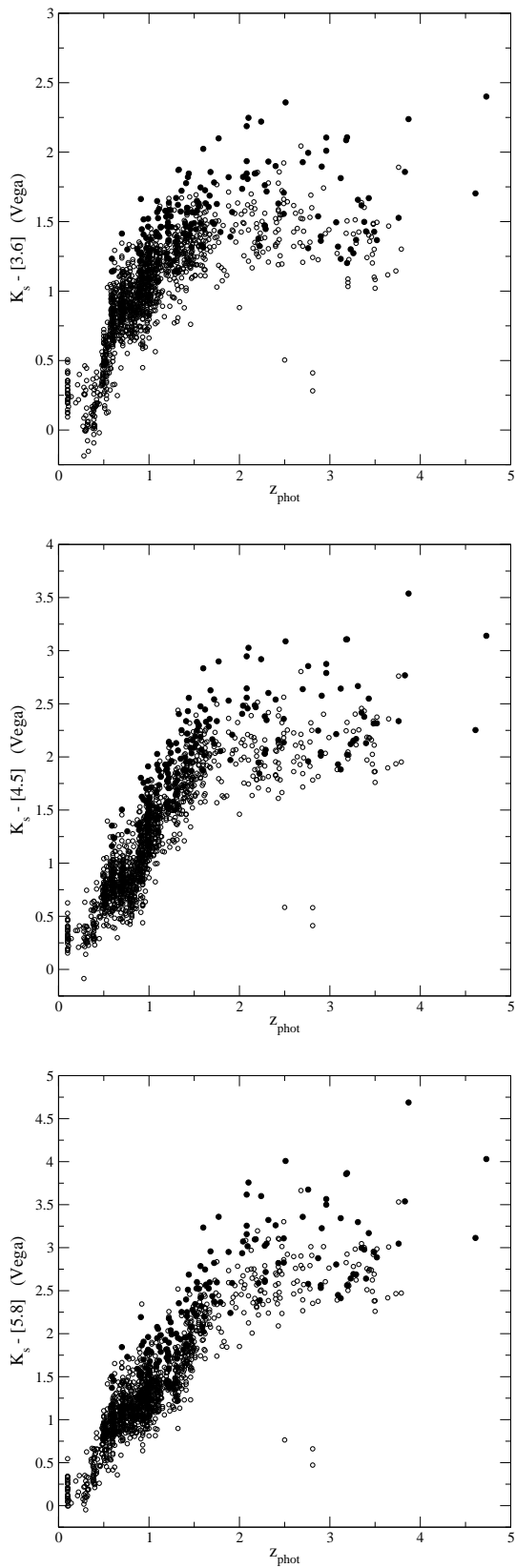}
\caption[]{\label{iraccol} Predicted IR colours of $\rm K_s$$\leq$22-selected sources as a function of estimated redshift. $[3.6]$, $[4.5]$ and $[5.8]$ indicate Spitzer/IRAC magnitudes at the channels $3.6 \mu m$, $4.5 \mu m$ and $5.8 \mu m$, respectively.}
\end{center}  
\end{figure}

   Figure \ref{iraccol} shows the predicted $\rm K_s - [3.6 \mu m]$ (a),  $\rm K_s - [4.5 \mu m]$ (b) and  $\rm K_s - [5.8 \mu m]$ (c) colours of $\rm K_s$-selected sources versus estimated redshift. The filled circles correspond to the ERGs and the open circles to all the other $\rm K_s$-selected galaxies. We observe that $\rm (I_{775}-K_s)$ selected ERGs do not have discriminating  near-IR colours, and a pure IR colour cut does not seem to select the same population as an $\rm (I_{775}-K_s)$ cut, as was pointed out by Wilson et al. (2004) (these authors compared a pure IR colour cut with an $\rm (R-K_s)$ cut). From the analysis of Spitzer/IRAC data, these authors also conclude that a $\rm K_s - [3.6 \mu m]$ red colour could be more effective to select high redshift ($z>1.3$) galaxies. The predicted IR colours for our deep sample of $\rm K_s$-selected galaxies leads us to  a similar conclusion. However, a $\rm K_s - [3.6 \mu m]$ colour cut  sufficiently red to avoid $z<1$ contaminants might miss some of the higher redshift objects. On the contrary, the $\rm K_s - [5.8 \mu m]$ colour appears as a rather better redshift indicator, and a red $\rm K_s - [5.8 \mu m]$ cut could more effectively separate the higher redshift galaxies. The relatively tight relation existing  between  the $\rm K_s - [5.8 \mu m]$ colour and redshift suggests that the three very blue sources with $z_{phot} \sim 2.5-3.0$ might have erroneous estimated redshifts.

\begin{table*}
\caption[]{\label{yancomp} $\rm K_s$$\leq$22-selected sources in the present sample which are among Yan et al. (2004) IRAC extremely red objects. The columns list the following: 1) our identification number; 2) Yan et al. identification; 3)-4) RA and DEC on the $\rm K_s$-band images; 5) our estimated redshift; 6) Yan et al. estimated redshift; 7) our measured $\rm K_s$-band   magnitude; 8) Yan et al. measured $\rm K_s$-band magnitude; 9)-10)-11) our predicted magnitudes in the IRAC channels $3.6 \mu m$, $4.5 \mu m$ and $5.8 \mu m$, respectively ; 12)-13)-14) magnitudes measured by Yan et al. in the IRAC channels $3.6 \mu m$, $4.5 \mu m$ and $5.8 \mu m$, respectively. All magnitudes are in the Vega system. To convert Yan et al. magnitudes quoted in the AB system to Vega magnitudes, we used: $\rm K_{s_{Vega}}=K_{s_{AB}}-1.84$, $\rm [3.6]_{Vega}=[3.6]_{AB}-2.79$,  $\rm [4.5]_{Vega}=[4.5]_{AB}-3.26$ and $\rm [5.8]_{Vega}=[5.8]_{AB}-3.73$. }
\begin{tabular}{rcccrlcrrrrlll}
\hline\hline
  id & id $\rm ^Y$ & RA(J2000) & DEC(J2000) &  $z_{phot}$ & $z_{p}$$\rm ^Y$ & $\rm K_s$ & $\rm K_s$$\rm ^Y$ & $\rm [3.6]$ & $\rm [4.5]$ &  $\rm [5.8]$ & $\rm [3.6]$$\rm ^Y$ & $\rm [4.5]$$\rm ^Y$ &  $\rm [5.8]$$\rm ^Y$ \\
\hline
1418 & 4 & 3:32:41.76 & -27:48:24.98 & 3.40 & 2.7 & 21.87  & 21.79 &  20.5 & 19.8 & 19.3 & 19.61 & 19.00 & 18.57 \\
1539 & 7 & 3:32:32.17 & -27:46:51.48 & 2.96 & 2.7 & 21.80  & 21.89  & 19.9 & 19.1 & 18.4 & 19.36 & 18.67 & 17.90 \\
1429 & 8 & 3:32:35.09 & -27:46:47.46 & 2.70 & 2.9 & 21.07  & 20.98 & 19.3 & 18.6 & 17.8 & 18.65 & 18.11 & 17.28 \\
146 &  9 & 3:32:39.17 & -27:48:32.33 & 2.71 & 2.8 & 21.18  & 21.03  & 19.8 & 19.2 & 18.6 & 19.14 & 18.69 & 18.01 \\
14115 & 13 & 3:32:39.12 & -27:47:51.45 & 1.92 & 1.9 & 21.10  & 20.80  & 19.5 & 18.9 & 18.5 & 19.19 & 18.77 & 18.14 \\
14181 & 16 & 3:32:35.72 & -27:46:38.73 & 1.42 & 2.4 & 21.32 & 21.50 & 19.9 & 19.3 & 19.2 & 19.35 & 18.84 & 18.27 \\ 
14106 & 17 & 3:32:33.67 & -27:47:51.07 & 1.70 & 1.6 & 20.54 & 20.55 & 19.2 & 18.5 & 18.3 & 18.92 & 18.50 & 18.22 \\
\hline
\hline
\end{tabular}
\end{table*}

    It is interesting to compare the predicted magnitudes for a subset of our sources with recent results obtained by direct photometric measurements on IRAC images. Yan et al. (2004) recently selected a sample of 17 IRAC objects with $f_{\nu}(3.6 \mu m)/f_{\nu}(z_{850})>20$ in the HUDF. We find that 8/17 of these objects are members of our $\rm K_s$$\leq$22-selected sample. The remaining objects are not in our sample either because they are fainter $\rm K_s$-band galaxies or because they lie outside of our field of view\footnote{The ISAAC field studied in this work only partially overlaps the HUDF.}.  1/8 of the common objects (identification number 1 in Yan et al. 2004) has a very uncertain estimated redshift in Yan et al. (2004) and is excluded from their refined sample. The remaining 7/8 common objects are listed in Table \ref{yancomp}. Our estimated redshifts  for the seven common sources are in good agreement with the Yan et al. estimations (median of $\delta z / (1+z) \approx 0.07 $). We note that, however, our predictions underestimate the observed magnitudes in the IRAC bands by a median of $\approx$ 0.5 mag. The comparison of the measured  $\rm K_s$ magnitudes in both cases shows that this systematic offset cannot be attributed to, for example, aperture effects. However, given that the sources listed in Table \ref{yancomp} have been deliberately selected by Yan et al. on the basis of their flux excess on the  IRAC bands, it is unsurprising (and reassuring) that, for this highly biased subset of sources, our predicted IRAC fluxes are systematically low. We anticipate a much better agreement of our predicted values with the observed IRAC magnitudes for the majority of our sample, when the GOODS/CDFS IRAC data become available.

\section{Summary and conclusions}

\parskip=0pt

 In this work we have presented the redshift distribution and resulting cosmological implications for a sample of 1663 $\rm K_s$-selected galaxies with $\rm K_s \le 22$ (Vega), and made an extensive comparison of the results with  those obtained in C04 for the R03 ERG sample in the same field. This is the deepest significant study to date of $\rm K_s$-selected galaxies and the role of ERGs within this population.

  We have studied the evolution of the rest-frame $\rm K_s$-band LF and concluded that there is no evidence for a decrease of its bright end up to at least redshift  $\langle z_{phot}\rangle = 2.5$. A similar conclusion has been reached for the ERGs in C04, because these objects account for most of the bright end of the $\rm K_s$-band LF at high redshifts. Also, from the comparison of both the total $\rm K_s$-band and the ERG LFs, we find that the latter appears to have a bell-like shape at redshift $z_{phot} \sim 1$, with little contribution to the faint end of the former.  This result is consistent with other previous works, which determined that red and blue galaxies dominate different parts of the total LF in the local Universe as well as higher redshifts. The total $\rm K_s$-band and the ERG LFs become progressively more similar with redshift, reaching a maximum at $\langle z_{phot}\rangle=2.0$. The limiting magnitude of our sample does not allow us to properly explore the faint end of the LF above this redshift.

  We also study the evolution of the massive systems present in our sample of  $\rm K_s$-selected galaxies. We determined that a significant fraction (up to 20\%-25\%) of the galaxies with mass $\rm M>1 \times 10^{11} M_\odot$ is in place before redshift $z_{phot} \sim 4$. However, the assembly of most of the remaining massive systems appears to have occurred at later epochs,  at redshift $z_{phot} \lsim 1.5$. ERGs at redshifts $z_{phot} \sim 1-2$ account for most of the massive galaxies which have plausibly evolved under passive evolution since very high redshifts, but 30\%-50\% of the galaxies with assembled stellar mass $\rm M>1 \times 10^{11} M_\odot$ at redshift $z_{phot} \sim 1-2$ are not ERGs. On the other hand, ERGs account for only a fraction of the massive systems present in the local Universe, indicating that, in any case,  passive evolution could not entirely explain the construction of all the massive galaxies.

  We found that the comoving number densities of galaxies with stellar mass $\rm M>1 \times 10^{11} M_\odot$ at redshifts $z_{phot} \sim  1-3$ is significantly smaller than the corresponding densities of dark matter haloes massive enough to host these systems. We suggest that this could be evidence for differential timescales in halo merging and galaxy collapse. At redshifts $z_{phot} \gsim 3$, however, the number of massive galaxies coincides with the corresponding number of haloes, probably indicating that galaxy collapse was a much faster and efficient process at very high redshift, virtually simultaneous with the virialisation of the haloes.

  From the analysis of the distribution of  the stellar mass budget at different redshifts, we conclude that between 45\% and 70\% of the stellar mass assembled at redshifts $3<z_{phot}<4$ is contained in galaxies with mass $\rm M>1 \times 10^{11} M_\odot$. These massive galaxies present at very  high redshift have presumably been formed from the early collapse of the high peaks of the density fluctuation field. Our study of the evolution of massive systems up to redshifts $z_{phot} \sim 4$ seems to confirm the existence of an evolutionary sequence in which most of the massive galaxies formed at very high redshift  become $\rm (I-K_s)$ selected ERGs by redshift $z_{phot} \sim 1$, and are the progenitors of the massive elliptical galaxies observed in local clusters.

   The follow up in the Spitzer/IRAC channels will be necessary to directly observe the rest-frame $\rm K_s$-band light of the high-redshift ($z>1$) galaxies in our sample. This will allow an independent test of our extrapolated stellar luminosities and masses at high redshift. Moreover, the study of complementary properties for $\rm K_s$-selected galaxies (e.g. morphology) should shed further light on the nature and evolution of these objects.


\section*{Acknowledgements}
This paper is based on observations made with the Advanced Camera for Surveys  and the Near Infrared Camera and Multi Object Spectrometer on board the Hubble Space Telescope operated by NASA/ESA and with the Infrared
Spectrometer and Array Camera on the `Antu' Very Large Telescope operated by the
European Southern Observatory in Cerro Paranal, Chile, and form part of the
publicly available GOODS datasets.  We thank the GOODS teams for providing
reduced data products. We thank Will Percival for providing the code to compute the comoving densities of dark matter haloes. 
We also thank Alan Heavens and Ben Panter for computing the evolution of the star formation rates based on SDSS data.

 KIC acknowledges funding from a POE-network studentship and the Overseas
 Research Scheme Award (ORS/2001014037). JSD, RJM and NDR acknowledge PPARC
 funding.


\bibliographystyle{mn2e}

\section*{References}

\bib Arnouts S. et al., 2001, A\&A, 379, 740

\bib Bell E.F., McIntosh D.H., Katz N., Weinberg M.D., 2003, ApJS, 149, 289

\bib Ben\'{\i}tez N., 2000, ApJ, 536, 571

\bib Bertin E., Arnouts S., 1996, A\&A, 117, 393

\bib Bolzonella M., Miralles J.-M., Pell\'o R., 2000, A\&A, 363, 476

\bib Bruzual A.G., Charlot S., 1993, ApJ, 405, 538

\bib Bruzual A.G., Charlot S., 2003, MNRAS, 344, 1000

\bib Calzetti D., Armus L., Bohlin R. C., Kinney A. L., Koornneef
J., Storchi-Bergmann T., 2000, ApJ, 533, 682

\bib Caputi K.I., Dunlop J.S., McLure R.J., Roche N.D., 2004, MNRAS, 353, 30 (astro-ph/0401047) (C04)

\bib Cimatti A. et al., 2002, A\&A, 381, L68

\bib  Cimatti A. et al., 2002b, A\&A, 392, 395

\bib Cole S. et al., 2001, MNRAS, 326, 255

\bib Daddi E. et al., 2000, A\&A, 361, 535

\bib Daddi E. et al., 2003, ApJ, 588, 50

\bib Dickinson M., Papovich C., Ferguson H., Budav\'ari T., 2003, ApJ, 587, 25

\bib Drory N., Bender R., Feulner G., Hopp U., Maraston C., Snigula J., Hill G.J., 2003, ApJ, 595, 698

\bib Drory N., Bender R., Feulner G., Hopp U., Maraston C., Snigula J., Hill G.J., 2004, ApJ, 608, 742

\bib Fazio G. G. et al., 2004, ApJS, 154, 10

\bib Fontana A. et al., 2003, ApJ, 594, L9

\bib Fontana A. et al., 2004, A\&A, 424, 23

\bib Franx M. et al., 2003, ApJ, 587, L79

\bib Giavalisco M., the GOODS team, 2004, ApJ, 600, L93

\bib Glazebrook K. et al., 2004,  Nature, 430, 181

\bib Heavens A., Panter B., Jim\'{e}nez R., Dunlop J., 2004, Nature, 428, 625

\bib Kauffmann G., Charlot S., 1998, MNRAS, 297, L23

\bib Kochanek C.S. et al., 2001, ApJ, 560, 566

\bib Kodama T. et al., 2004, MNRAS, 350, 1005

\bib Le F\`evre O. et al., 2004, A\&A, 428, 1043

\bib Lilly S.J., Tresse L., Hammer F., Crampton D., Le F\`evre O., 1995, ApJ, 455, 108  

\bib Mo H.J., Jing Y.P, White S.D.M., 1997, MNRAS, 284, 189

\bib Moustakas L. et al., 2004, ApJ, 600, L131 (astro-ph/0309187)

\bib Poggianti B.M., 1997, A\&AS, 122, 399

\bib Pozzetti L. et al., 2003, A\&A, 402, 837

\bib Roche N.D., Almaini O., Dunlop J.S., Ivison R.J., Willott C.J., 2002, MNRAS, 337, 1282

\bib Roche N.D., Dunlop J.S., Almaini O., 2003, MNRAS, 346, 803 
(astro-ph/0303206) (R03)

\bib Rudnick G. et al., 2003, ApJ, 599, 847

\bib Somerville R.S., Primack J.R., 1999, MNRAS, 310, 1087

\bib Somerville R.S., Lee K., Ferguson H.C., Gardner J.P., Moustakas L.A., Giavalisco M., 2004, ApJ, 600, L171 (astro-ph/0309071)

\bib Szokoly et al., 2004, ApJS, 155, 271 

\bib Vanzella E. et al., 2004, The Messenger, 118, 45

\bib Willott C.J., Rawlings S., Jarvis M.J., Blundell K.M.,
2003, MNRAS, 339, 173

\bib Wilson G. et al., 2004, ApJS, 154, 107

\bib Yan H. et al., 2004, ApJ accepted (astro-ph/0408070)

\end{document}